\documentclass[aps, prd, twocolumn, showpacs, superscriptaddress, nofootinbib,floatfix]{revtex4-2}
\usepackage{amsmath,mathrsfs,amsbsy,color,graphicx,bm,amsthm,amsfonts}
\usepackage{bbm}
\usepackage{bm}
\usepackage{times}
\usepackage{units}
\usepackage{dcolumn}
\usepackage{graphicx}
\usepackage{epsfig}
\usepackage{epstopdf}
\usepackage[colorlinks,linkcolor=red,anchorcolor=green,citecolor=blue,CJKbookmarks=True]{hyperref}
\DeclareMathSymbol{\shortminus}{\mathbin}{AMSa}{"39}
\usepackage{mathrsfs}
\usepackage{braket}
\usepackage{amssymb}
\usepackage{txfonts}
\usepackage{enumitem}
\usepackage{multirow}
\graphicspath{{fihures/},{pics/}} \usepackage{subfigure} 
\usepackage{microtype}
\usepackage[cmyk]{xcolor} 
\usepackage{placeins}

\usepackage{orcidlink}

\begin{document}
\title{Scrambling-Enhanced Quantum Battery Charging in Black Hole Analogues}
	\author{Zhilong Liu~\orcidlink{0009-0000-0353-5113}}
	\affiliation{Department of Physics, Key Laboratory of Low Dimensional Quantum Structures and Quantum Control of Ministry of Education, and Synergetic Innovation Center for Quantum Effects and Applications, Hunan Normal
		University, Changsha, Hunan 410081, P. R. China}
	\affiliation{ Institute of Interdisciplinary Studies, Hunan Normal University, Changsha, Hunan 410081, P. R. China}
		
	\author{Ying Li~}
	\affiliation{Department of Physics, Key Laboratory of Low Dimensional Quantum Structures and Quantum Control of Ministry of Education, and Synergetic Innovation Center for Quantum Effects and Applications, Hunan Normal
		University, Changsha, Hunan 410081, P. R. China}
	\affiliation{ Institute of Interdisciplinary Studies, Hunan Normal University, Changsha, Hunan 410081, P. R. China}
	
	\author{Zehua Tian}
	\email{tzh@hznu.edu.cn (Corresponding author)}\affiliation{School of Physics, Hangzhou Normal University, Hangzhou, Zhejiang 311121, China}
	
	\author{Jieci Wang~\orcidlink{0000-0001-5072-3096}}
	\email{jcwang@hunnu.edu.cn (Corresponding author)}\affiliation{Department of Physics, Key Laboratory of Low Dimensional Quantum Structures and Quantum Control of Ministry of Education, and Synergetic Innovation Center for Quantum Effects and Applications, Hunan Normal
		University, Changsha, Hunan 410081, P. R. China}
		\affiliation{ Institute of Interdisciplinary Studies, Hunan Normal University, Changsha, Hunan 410081, P. R. China}


\begin{abstract}Black holes constitute nature's fastest quantum information scramblers, a phenomenon captured by gravitational analogue systems such as position-dependent XY spin chains. In these models, scrambling dynamics are governed exclusively by the hopping interactions profile, independent of system size. Utilizing such curved spacetime analogues as quantum batteries, we explore how the  black hole scrambling affects charging via controlled quenches of preset scrambling parameters.
Our analysis reveals that the intentionally engineered difference between post-quench and pre-quench scrambling parameters could significantly enhance both maximum stored energy $E_{\max}$ and peak charging power $P_{\max}$ in the quench charging protocol. Furthermore, the peaks of extractable work and stored energy coincide. This is because the system's evolution under a weak perturbation remains close to the ground state, resulting in a passive state energy nearly identical to the ground state energy. The optimal charging time $\tau_*$ exhibits negligible dependence on the preset initial horizon parameter $x_{h0}$, while decreasing monotonically with increasing quench horizon parameter $x_{ht}$. This temporal compression confines high-power operation to regimes with strong post-quench scrambling $x_{ht} > x_{h0}$, demonstrating accelerated charging mediated by spacetime-mimicking scrambling dynamics.
\end{abstract}
\pacs{~}
\maketitle

\section{Introduction}
    The generation and storage of energy have historically been intricately linked to technological revolutions. The development of steam power and electricity has catalyzed two significant technological revolutions and laid the foundation for the third technological revolution. 
    With the rapid advancements in quantum technology and its applications across diverse fields—such as quantum communication \cite{Pirandola:2017kzd},
    quantum sensing \cite{Liu:2021frj,Zhang:2020skj,Degen:2016pxo}, quantum clock \cite{Wang:2019aqr,Zhang:2018atw,Wang:2015yma}, and quantum cryptography \cite{Garms:2024hip,Joseph:2022kac,Portmann:2021byc} —new applications are being actively explored as the capabilities to manipulate and measure quantum systems continue to improve. These technological capabilities are, in turn, paving the way for advanced theoretical research—not only in pushing the frontiers of quantum theory itself but also in exploring its connections with other fundamental frameworks, such as the interplay between quantum theory and general relativity ~\cite{Aziz:2025ypo,Marletto:2024ltk,article,Feng:2025djf}. Recently, attention has turned to quantum thermodynamics to investigate the potential for developing energy technologies within the quantum domain \cite{Kieu:2004fbx,Hovhannisyan:2013pqk,e18040124,Uzdin:2015gcs}. Notably, the concept of ``quantum batteries''---small quantum systems designed to temporarily store energy for subsequent use---has sparked significant research interest. This concept was first proposed by Alicki and Fannes \cite{Alicki:2013cwy}, who demonstrated that collective charging through entanglement operations outperforms parallel charging in terms of work extraction, which has attracted a series of studies focusing on optimizing charging performance through quantum effects \cite{Andolina:2024iet,Hu:2025sia,Goold:2015cqg,PhysRevLett.118.150601,Julia-Farre:2020jke}. 
   
    Since Ferraro introduced a quantum battery model based on the Dicke model that achieves superextensive power scaling \cite{Ferraro:2018tum}, various theoretical frameworks of quantum batteries have been proposed, including spin-chain model of a many-body quantum battery \cite{Le:2018fet,Grazi:2025tlx,K:2024sik,Grazi:2024kyr,Grazi:2025gxl}, Sachdev-Ye-Kitaev (SYK) models \cite{Divi:2024mup,Rossini:2019nfu,Rosa:2019jin,Romero:2024wgt}, cavity-QED architectures \cite{Wang:2024ogv,Hu:2025aod}, hybrid cavity-spin chain systems \cite{PhysRevA.106.032212,Zhao:2025iki,Sun2024CavityHeisenbergSC,Zhang:2024bva,Zhang:2024hvm}, continuous variable quantum batteries~\cite{Konar:2022myn,Downing:2024qzh, Downing:2025hfg} ,and open quantum battery models in curved spacetime \cite{Tian:2024wby, 2025arXiv250607568L, Hao:2023ndo,Liu:2025bzv,Chen:2025xaf,Xie:2024hwg,Mukherjee2024EnhancementOA}. Comprehensive reviews of these quantum battery models are available in Ref.~\cite{Campaioli:2023ndh}. Recently, experimental progress has also been made in realizing quantum batteries \cite{Hu:2021klf,Joshi:2021aee}. Research on charging performance metrics such as optimal charging time, maximum power, and extractable work is crucial for enhancing quantum battery efficiency. Efficient and rapid charging, along with stable energy storage, remain key areas of focus.

    In this context, quantum chaos, due to its ability to rapidly scramble information within the system, may influence the rate at which the system populates higher energy states, thereby exerting a positive effect on the charging process of quantum batteries.
    The SYK model has garnered significant attention within the field as an object of study for chaotic systems \cite{Kobrin:2020xms,Gyhm:2023lhb,Rosa:2019jin}. 
    A recent study utilizing the SYK model examined the quantum chaotic scrambling effects on the dynamics of quantum battery charging, revealing that while scrambling has the potential to expedite the charging process, it does not universally diminish charging times \cite{Romero:2024wgt}. The authors suggest that the overall impact on charging efficiency may be critically contingent upon the energy resources available within the charging Hamiltonian. 
    Studies by Gyhm and Fischer \cite{Gyhm:2023lhb} suggest that the chaotic final state of the SYK model does not provide a quantum advantage; however, scrambling behavior in early stages does enhance performance. Furthermore, analysis of the work by Erdman \emph{et al}. \cite{Erdman:2022oyp} on parameterized machine learning optimization of the Dicke-model quantum battery indicates that the machine learning process does not lead to further improvements in the early-stage dynamics. These collective findings suggest that, during the early scrambling period, pure scrambling behavior could have a positive effect on the charging process.
    Additionally, as extreme celestial objects in nature, black holes serve as the fastest information scramblers \cite{Maldacena:2015waa,Shenker:2013pqa}. Yang \emph{et al}. investigated the simulation of curved spacetime effects using an isotropic XY model with position-dependent hopping interactions and numerically studied chaotic behavior analogous to that of black holes within the AdS/CFT framework \cite{Yang:2019kbb,Liu:2024wqj}. 
    Shi \emph{et al}. constructed an analog quantum simulation platform using a chain of superconducting transmon qubits with interactions mediated by transmon-type tunable couplers, providing experimental demonstration of curved spacetime simulations \cite{Shi:2021nkx}. 
    Besides, similar models have been designed to study various relativistic physics in curved spacetime \cite{Houck:2012hbf,Kinoshita:2024ahu,Deger:2022qob,Wang:2020ypl,Koke:2016etw,Toga2025FastSI,Gong2025DigitQS,Alkac:2025hrv,Li2024SimulatingTS,Horner:2022sei,Kinoshita2025SpinSA,Daniel:2024boy,Jaiswal:2025euo,Moghaddam2025SyntheticHA,Daniel2025QuantumTB}. 
    The chaotic effects in these black hole analogues are determined exclusively by intrinsic parameters, namely the distribution of nearest-neighbor hopping interactions in the XY chain. This parameter-dependence contrasts with system-size-dependent chaotic behavior captured in other models such as the SYK system. This enables a more focused investigation into how chaos influences quantum battery charging by minimizing the impact of extraneous variables.

    By capitalizing on the characteristic that the scrambling intensity of this analogue system is independent of its size, we examine how this chaotic scrambling influences the charging process of quantum battery based on the gravitational analogue system. The quantum battery implementation employs an isotropic XY chain with position-dependent hopping interactions and a scrambling parameter quench protocol. Motivated by the rapid information scrambling characteristic of chaotic dynamics, we quantify its impact on charging efficiency through three key metrics: maximum stored energy \(E_{\text{max}}\), maximum charging power \(P_{\text{max}}\), optimal charging time \(\tau_{*}\).
     Systematic analysis reveals how scrambling intensity modulates these performance indicators. 
      \begin{figure}[t]
         \centering{\includegraphics[width=0.98\linewidth]{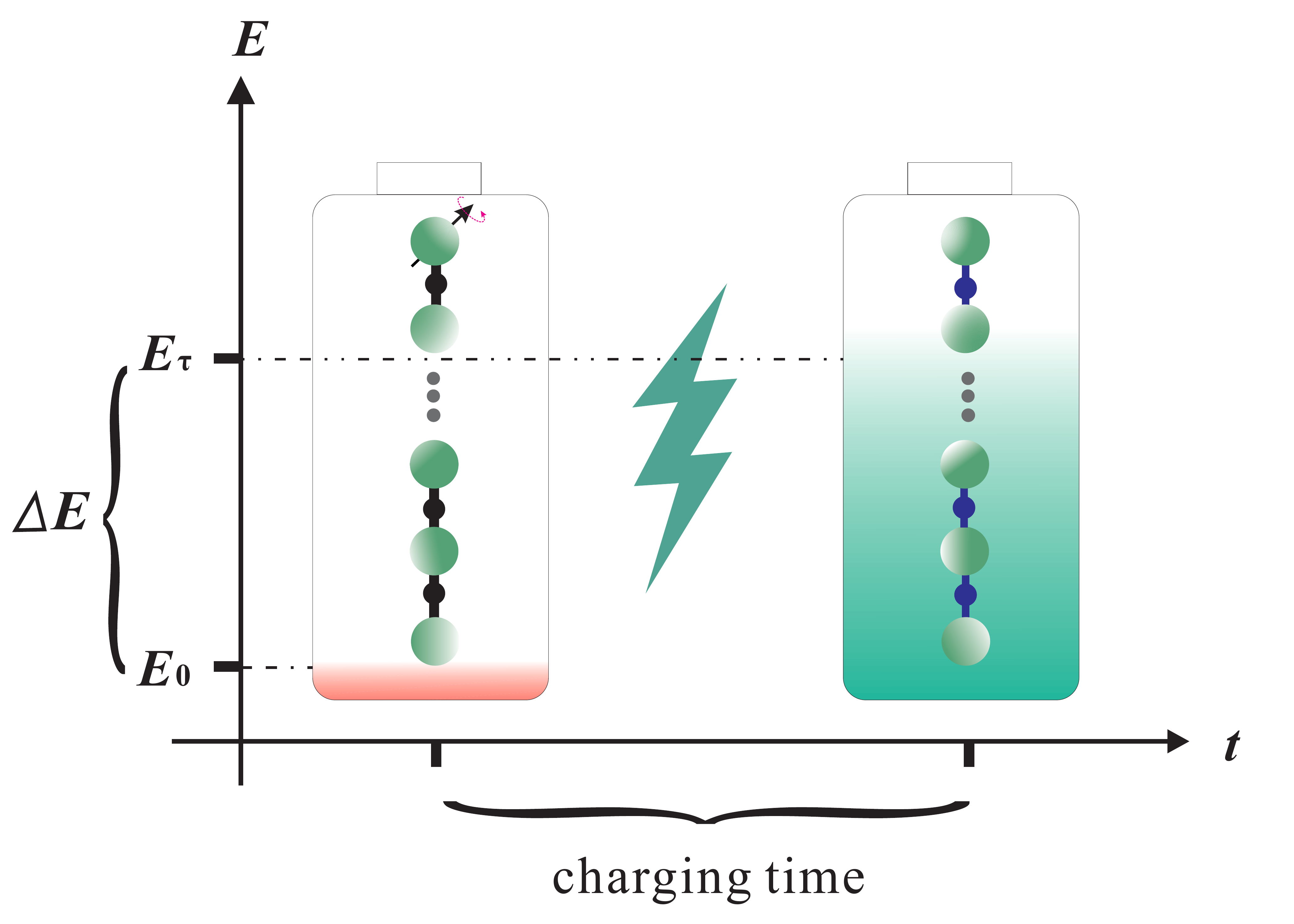}
         \caption{ Schematic diagram of the charging process: The quantum battery employs a nearest-neighbor hopping XY spin chain architecture. Green spheres represent individual spin qubits, interconnected by tunable couplers. Distinct coupling distribution groups are represented by unique colors.Notably, the coupling distribution during the charging process differs from that observed in the non-charging process.}}
     \end{figure}
    
	The paper is organized as follows. Section \ref{Sec_II} introduces the quantum battery model based on an isotropic XY spin chain with position-dependent hopping interactions and discusses its chaotic scrambling behavior. Section \ref{sec_experi} provides a detailed analysis of the experimental implementation. Numerical results on the influence of scrambling effects on the battery's charging dynamics are presented in Section \ref{Sec_III}. The impact of system size and bandwidth regularization is then discussed in Section \ref{Sec_IV}. Finally, concluding remarks are provided in Section \ref{Sec_V}.
\section{Theoretical Model and Experimental Prospects}
\subsection{Quantum battery model  and scrambling}\label{Sec_II}
    In this section, we introduce the fundamental components of the gravitational analog system, focusing on its Hamiltonian formulation and chaotic dynamics, followed by a concise overview of the quantum battery model derived from this system. Specifically, we consider a 1+1 dimensional black hole spacetime with the line element    
     \begin{equation}
        \mathrm{d}s^2=-f(x)\mathrm{d}t^2=\frac{1}{f(x)}\mathrm{d}x^2,
    \end{equation}
    where the signature is set (-,+). Then the massless Dirac field in this spacetime is equivalent to an isotropic one-dimensional XY spin chain with position-dependent hopping interactions, governed by the Hamiltonian \cite{Yang:2019kbb}:
        \begin{equation}
            H_{XY}=\sum_{n=1}^L \left[-\kappa_n(\sigma^x_n\sigma^x_{n-1}+\sigma^y_n\sigma^y_{n-1})-\mu \sigma_n^z\right]\label{eq_XY},
        \end{equation}
    where $\sigma^i$ represents Pauli matrices, the position-dependent nearest-neighbor hopping interaction strengths are defined as \(\kappa_n = f\left[\left(n - \frac{1}{2}\right)d\right]/4d\), where \(f(n)\) represents the spatially discretized version of the metric function \(f(x)\), and \(d\) denotes the discretization interval. Furthermore, \(\mu\) and \(L\) denote the on-site potential and system size, respectively. The geometric information of the curved spacetime, represented by the metric function $f(x)$, is encoded in the spatial distribution of the coupling strengths. Through the Jordan-Wigner transformation \cite{Barouch:1970ryz,Barouch:1971ywx}:
        \begin{subequations}
            \begin{align}
                \sigma_n^+ &= e^{i\pi \sum_{j<n} \hat{c}_j^\dagger \hat{c}_j}\hat{c}_n^\dagger, \label{eq_JWa}\\
                \sigma_n^- &= \hat{c}_n e^{-i\pi \sum_{j<n} \hat{c}_j^\dagger \hat{c}_j}, \\
                \sigma_n^z &= 1 - \hat{c}_n^\dagger\hat{c}_n, \label{eq_JWc}
            \end{align}
        \end{subequations}
    \begin{figure}[b]
            \centering{\includegraphics[width=0.98\linewidth]{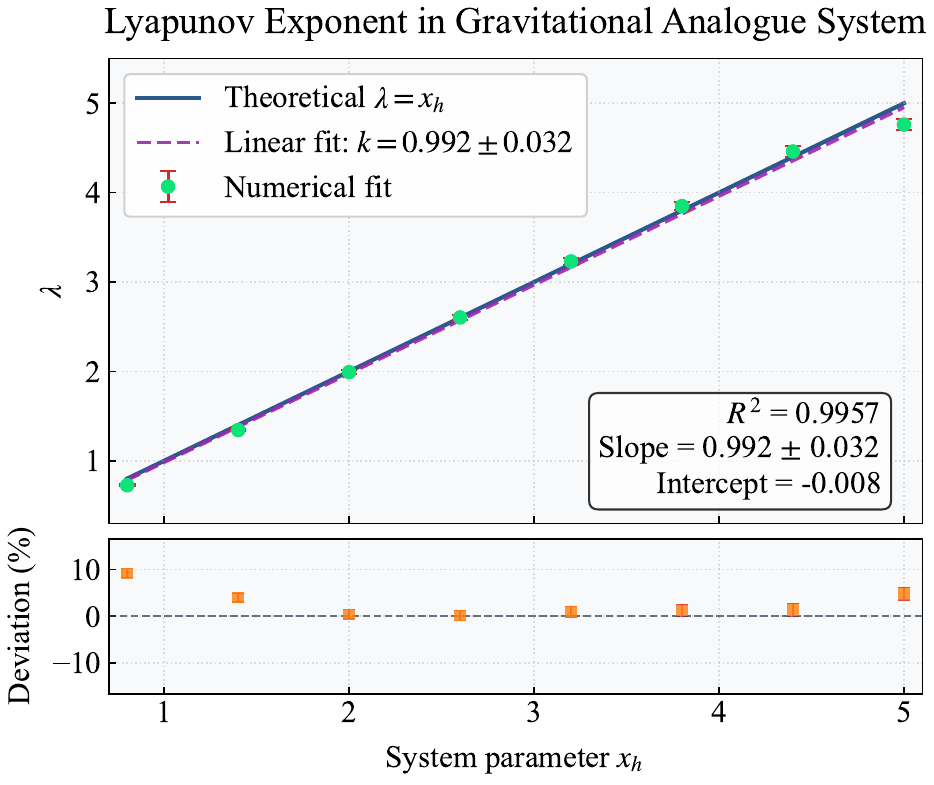}
            \caption{Variation of the fitted Lyapunov exponent $\lambda$ as a function of the horizon parameter $x_h$. We select metric function $f(x)=x^2(1-x_h/x)$ such that the Hawking temperature is given by $T=x_h/(4\pi)$. The theoretical chaos bound is represented by the blue curve, while the numerical data is depicted as green dots with red error bars. The size of system set L=250.}\label{fig_2}
            }
        \end{figure}
    with
        \begin{align}
                \sigma_n^+ =\frac{1}{2}(\sigma_n^x +i \sigma_n^y),    \qquad   \sigma_n^- = \frac{1}{2}( \sigma_n^x -i \sigma_n^y),
        \end{align}
    this system can be transformed into a free Hubbard model with site-dependent hopping, governed by the Hamiltonian 
         \begin{equation}
             H=\sum_{n}^L \left[-\kappa_n(\hat{c}_n^\dagger\hat{c}_{n-1}+\hat{c}_{n-1}^\dagger\hat{c}_{n})+\frac{\mu}{2} \hat{c}_n^\dagger\hat{c}_{n}\right].\label{eq_hub}
         \end{equation}
    Here, \(\hat{c}^\dagger_n\) and \(\hat{c}_n\) denote fermionic ladder operators that satisfy the anti-commutation relations: \(\{\hat{c}_m, \hat{c}^\dagger_n\} = \delta_{mn}\) and \(\{\hat{c}_m, \hat{c}_n\} = \{\hat{c}^\dagger_m, \hat{c}^\dagger_n\} = 0\). The Hamiltonian of the analogue gravity system can inherit most properties of curved spacetime, including, e.g., the chaotic behavior characteristic of black hole systems. To illustrate this, we utilize the so called out-of-time-order correlation (OTOC),  to diagnose the chaos. The OTOC is defined as follows \cite{Maldacena:2015waa,Garcia-Mata:2022voo,PhysRevLett.117.091602}:
        \begin{eqnarray}
            C(t)=-\langle [W(t), V(0)]^2 \rangle,
        \end{eqnarray}
    where $W$ and $V$ are local Hermitian operators that may be either the same or different. For a chaotic system, this correlation satisfies the relation $C(t) \propto e^{2\lambda_Lt} $ with  $\lambda_L$ being the Lyapunov exponent \cite{Maldacena:2015waa,Garcia-Mata:2022voo,PhysRevLett.117.091602}. It was conjectured that the Lyapunov exponent $\lambda_L$ 
    characterizing chaos in thermal quantum many-body systems has a rigorous bound, $\lambda_L\leq2\pi T/\hbar$ \cite{Maldacena:2015waa}, where $T$ is the temperature of system, as
    established by Maldacena, Shenker, and Stanford (MSS). 
    Furthermore, this MSS bound is saturated when quantum system is an exact holographic dual to a black hole \cite{Maldacena:2015waa}. Recently, an experimental implementation has been proposed to verify whether the MSS bound can be identically saturated in a concrete analogue-gravity system using a trapped ion \cite{Tian:2020bze}.
   
     We consider an asymptotically AdS\(_2\) black hole spacetime with the metric function 
    \begin{equation}
        f(x) = x^2 (1 - x_h/x),
    \end{equation}
    where \(x_h\) denotes the event horizon radius. The Hawking temperature is given by \(T = x_h/(4\pi)\). Since black holes saturate the chaos bound as the fastest scramblers in nature, the Lyapunov exponent should obey the relation~\cite{Maldacena:2015waa,Yang:2019kbb}
        \begin{equation}
            \lambda_L = 2\pi T = x_h/2, 
        \end{equation}
    indicating exclusive dependence on the event horizon parameter. 

    In Fig.~\ref{fig_2}, we numerically compute the OTOC \(C(t)\) for varying horizon radii \(x_h\), extracting the corresponding exponents through curve fitting: \(\lambda_{\text{fit}} = a \cdot x_h + b\). We compare these results with the theoretical chaos bound curve (blue line): \(\lambda_{\text{fit}} = 2\lambda_L = x_h\). These numerical values are represented as green dots with red error bars in the figure. The fitted exponents show excellent agreement with the theoretical chaos bound, yielding a slope of \(0.992 \pm 0.032\) from linear regression. This demonstrates that chaotic behavior in this system depends solely on the horizon parameter and scales linearly with it. Given that the information of the curved spacetime is encoded in a spatially discrete manner within the nearest-neighbor hopping interactions, the chaotic scrambling behavior induced by the black hole spacetime in the gravitational analogue system is solely determined by the spatial distribution of these hopping interactions $\kappa_n$.    
    
    Next, we investigate the influence of chaotic effects in this analogue black hole spacetime on the quantum battery charging process through a parameter quench protocol within the nearest-neighbor gravitational analog system. The Hamiltonian for the entire system can be expressed as:
        \begin{align}
                H_{sys}(t) =& H_0+ \lambda(t) (H_1-H_0) \nonumber \\
                =&\sum_{n=1}^L \left[-\kappa_n(t)(\sigma^x_n\sigma^x_{n+1}+\sigma^y_n\sigma^y_{n+1})+\mu \sigma_n^z\right],
        \end{align}
     with the hopping interaction parameters defined as 
        \begin{equation}\label{eq_kappa}
            \kappa_n(t) = \kappa^{(0)}_n +  \lambda(t)(\kappa^{(1)}_n-\kappa^{(0)}_n) ,
        \end{equation}
    where $\lambda(t) = \Theta(\tau - t)\Theta(t)$ with $\Theta$ denoting the Heaviside step function, such that the quench occurs for $t \in (0,\tau]$. Here, the nearest-neighbor hopping distribution $\kappa_n$ depends solely on the scrambling parameter $x_{\mathrm{h}}$ (the horizon size that determines scrambling intensity). For clarity, distinct scrambling parameters $x_{\mathrm{h}}$ will denote different hopping distributions $\kappa_n$: the initial distribution $\kappa^{(0)}_n$ corresponds to $x_{\mathrm{h0}}$ (pre-quench), while the quench distribution corresponds to $x_{\mathrm{ht}}$. The stored energy during quench charging is thus given by
        \begin{equation}
            \Delta E = \langle H_0 \rangle_t - \langle H_0 \rangle_0,
        \end{equation}
    where \(\langle \cdot \rangle_t \equiv \langle \psi(t) | \cdot | \psi(t) \rangle\) denotes the quantum expectation value at time \(t\), and the initial state \(\ket{\psi(0)} = \ket{0}\) corresponds to the ground state of the unquenched system with hopping parameters \(\kappa_n = \kappa_n^{(0)}\).
    To enhance visualization, we normalize the stored energy as \(\Delta E \rightarrow \Delta E / E_b\), where the energy bandwidth \(E_b\) is defined as \(E_b = E_p - E_0\), with \(E_p\) and \(E_0\) representing the maximum and minimum eigenvalues of the system's Hamiltonian in the non-charging state, respectively.
    Since chaotic dynamics can rapidly scramble the local perturbation or information of the system to the whole system or all the degree of freedom of the system, we expect the primary advantage of chaotic effects to be manifest in enhanced charging performance in the charging protocol.
    To quantitatively characterize the scrambling dynamics during charging, we introduce the following performance metrics: 
    \begin{itemize}
        \item The maximum stored energy \(E_{\text{max}} \equiv \max_t \Delta E(t)\),
        \item The maximum charging power \(P_{\text{max}} \equiv \max_\tau {P}(\tau)\), where the average power \({P}(\tau) \equiv \Delta E(\tau)/\tau\),
        \item The optimal charging time \(\tau_{*}\) corresponding to \(P_{\text{max}}\), defined as \({P}(\tau_*) \equiv \max_\tau {P}(\tau)\).
    \end{itemize}   
    
\subsection{Experimental Feasibility and Proposed Implementation}\label{sec_experi}
	The experimental realization of our proposed quantum battery model, which maps to an XY spin chain with position-dependent coupling, is feasible on several advanced quantum simulation platforms, most notably superconducting quantum processors~\cite{Wei:2016jjg,Wang:2025ade,Li:2025kje,Zhao:2025jcc,Liu:2025tys,Alam:2025ofn}. These systems are particularly suitable due to the high degree of controllability over qubit frequencies and inter-qubit couplings. A practical implementation can be achieved using a chain of tunable transmon qubits with dedicated tunable couplers. The system Hamiltonian for a chain of $n$ qubits and $n-1$ couplers is~\cite{Yan:2018mli,Shi:2021nkx}
	
	\begin{equation}  
		\hat{H}= \hat{H}_Q+\hat{H}_C+\hat{H}_{Q-Q}+\hat{H}_{Q-C}+\hat{H}_{C-C},
	\end{equation}  
	wherein  
	\begin{eqnarray}  
		\hat{H}_{Q}=\sum_{j=1}^n\omega_{q_j}\hat{b}^\dagger_{q_j}\hat{b}_{q_j}+\frac{\alpha_{q_j}}{2}\hat{b}^\dagger_{q_j}\hat{b}^\dagger_{q_j}\hat{b}_{q_j}\hat{b}_{q_j}, \\
		\hat{H}_{C}=\sum_{j=1}^{n-1}\omega_{c_j}\hat{b}^\dagger_{c_j}\hat{b}_{c_j}+\frac{\alpha_{c_j}}{2}\hat{b}^\dagger_{c_j}\hat{b}^\dagger_{c_j}\hat{b}_{c_j}\hat{b}_{c_j}, \\
		\hat{H}_{Q-Q(C-C)}=\sum_{j=1}^{n-1}g_{q_j(c_j),q_{j+1}(c_{j+1})}\left(\hat{b}^\dagger_{q_j(c_j)}\hat{b}_{q_j(c_j)}+h.c\right),\\
		\hat{H}_{Q-C}=\sum_{j=1}^{n-1}\left(g_{q_j,(c_{j})}\hat{b}^\dagger_{q_j}\hat{b}_{c_j}+g_{q_{j+1},(c_{j})}\hat{b}^\dagger_{q_{j+1}}\hat{b}_{c_j}+h.c\right),
	\end{eqnarray}  
	where \( \omega_j \) and \(\alpha_j\) correspond to the frequencies and anharmonicities of the respective transmon qubits and \(g_{q_{j+1}(c_{j+1}),q_j(c_j)}\) represents the strength of the nearest-neighbor exchange interaction. By biasing the couplers and qubits at specific working points (e.g., via external magnetic fluxes for Josephson junction-based devices), one can create a large detuning condition. Applying the Schrieffer-Wolff transformation under this regime~\cite{Bravyi:2011fda}
	\begin{equation}
		\hat{U}=exp\left\{\sum_{j=1}^{n-1}\left(\frac{g_{q_j,c_j}}{\omega_{q_j}-\omega_{c_j}}\hat{b}^\dagger_{q_j}\hat{b}_{c_j}+\frac{g_{q_{j+1},c_j}}{\omega_{q_{j+1}}-\omega_{c_j}}\hat{b}^\dagger_{q_{j+1}}\hat{b}_{c_j}+h.c\right)\right\},
	\end{equation}
	which yields an effective Hamiltonian that is exactly our target XY chain model in Eq.~(\ref{eq_XY}). Subsequently, invoking the Jordan-Wigner transformation Eqs.~(\ref{eq_JWa})--(\ref{eq_JWc}), we obtain a free Hubbard model with site-dependent hopping, governed by the Hamiltonian in Eq.~(\ref{eq_hub}).
	Consequently, the crucial coupling distribution \( \{\kappa_n\} \) prescribed by our model is directly implemented through the designed capacitances of the circuit and can be dynamically tuned by programming the magnetic fluxes on the tunable couplers. Finally, the system's dynamics during the charging process can be tracked using quantum state tomography techniques, thereby enabling precise measurement of the energy storage and power output to verify the predicted quantum advantage.
\section{Scrambling in charging protocol}\label{Sec_III}
        \begin{figure}[b]
                \centering{\includegraphics[width=0.98\linewidth]{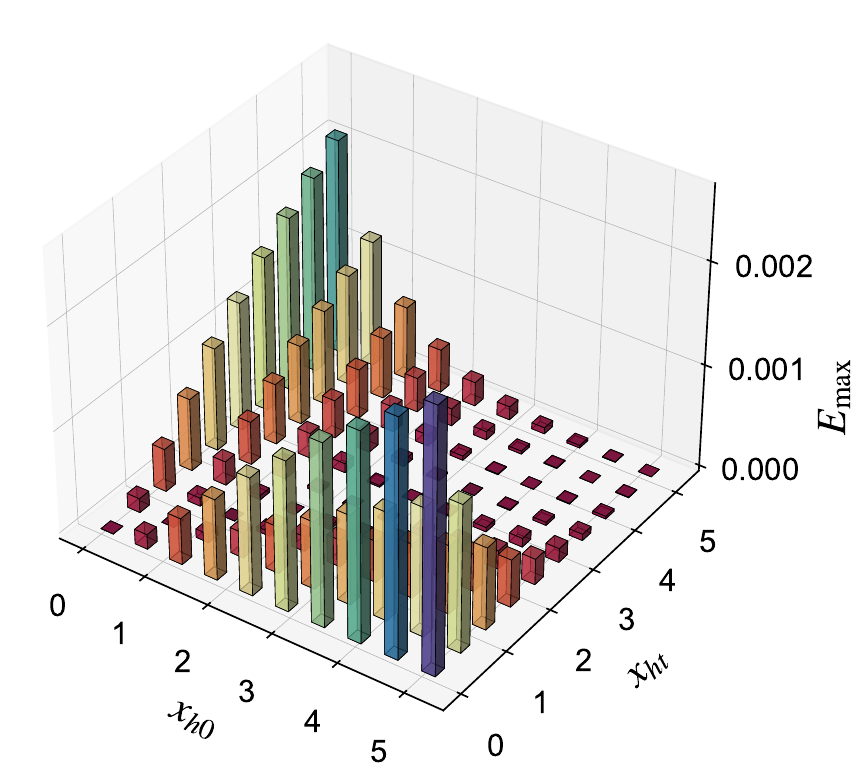}
                \caption{The plot of maximum stored energy \(E_{\text{max}}\) versus initial scrambling parameter $x_{h0}$ and quench scrambling parameter $x_{ht}$, with values represented by bar height or color (red to blue). The system size is set to \(L = 250\).}\label{fig_3}
                }
        \end{figure}
         \begin{figure}[b]
        	\centering{\includegraphics[width=0.98\linewidth]{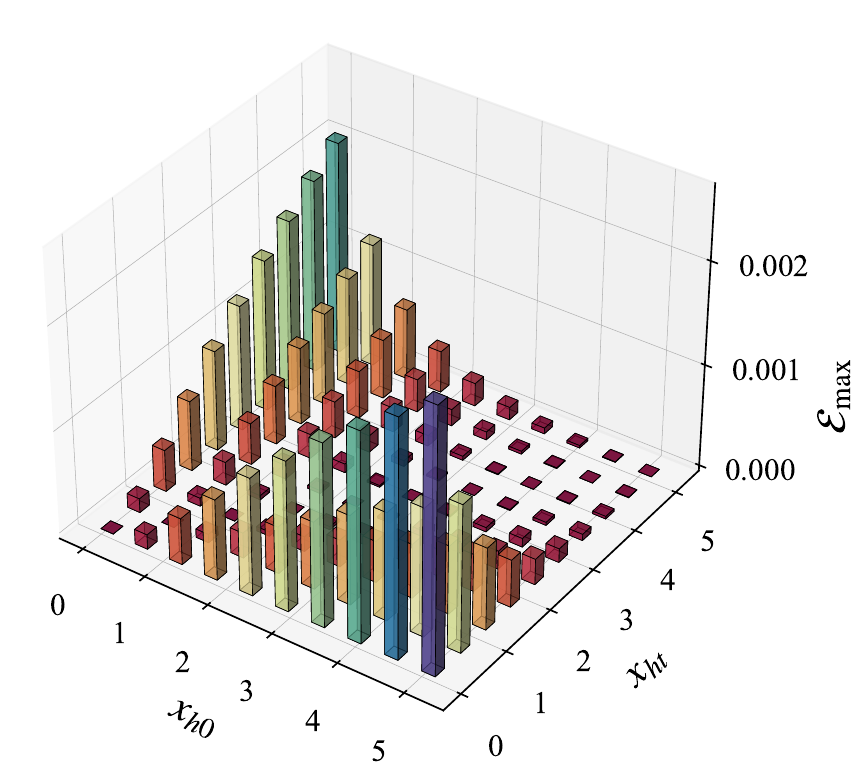}
        		\caption{The plot of maximum ergotropy \(\mathcal{E}_{\text{max}}\) versus initial scrambling parameter $x_{h0}$ and quench scrambling parameter $x_{ht}$, with values represented by bar height or color (red to blue). The system size is set to \(L = 250\).}\label{fig_ergo}
        	}
        \end{figure}
    In this section, we implement the quantum battery protocol within the black hole analog system described previously. 
    The charging process is induced by quenching the scrambling parameter \(x_h\), where the profile of position-dependent hopping strengths is governed by Eq.~(\ref{eq_kappa}). 
    Prior to charging, the XY chain resides in its ground state with initial distribution of hopping parameters given by \(\kappa_n^{(0)}\) and scrambling parameter is denoted as \(x_{h0}\). 
    At \(t = 0\), the quench instantaneously switches the hopping strengths profile to \( \kappa_n^{(1)}\), corresponding to a new scrambling parameter \(x_{ht}\). The non-commutativity of the evolution and initial Hamiltonians, \([H_t, H_0] \neq 0\), drives the charging process of the system. The charging concludes at time \(\tau\) when the hopping parameters are restored to their initial configuration, after which the system energy remains constant. The maximum stored energy during this process is determined via \(E_{\text{max}} \equiv \max_t \Delta E(t)\).

    As shown in Fig.~\ref{fig_3}, we numerically compute the maximum stored energy \(E_{\text{max}}\) for various combinations of initial and quenched scrambling parameters \(x_{h0}\) and \(x_{ht}\), presenting the results in a three-dimensional plot. To provide an intuitive and accessible visualization, the energy value is encoded in two equivalent ways: primarily by the bar's height, and redundantly by its color intensity.

    The characteristic red valley along the diagonal (\(x_{h0} = x_{ht}\)) corresponds to vanishing energy storage, where the commutativity \([H_0, H_t] = 0\) maintains the system in its ground state. This diagonal divides the parameter space into two distinct regions. 
    As the parameter difference \(|\Delta x_h| = |x_{ht} - x_{h0}|\) increases, \(E_{\text{max}}\) grows monotonically due to enhanced non-commutativity between the initial and charging Hamiltonians. In the region where \(x_{ht} > x_{h0}\) (stronger post-quench scrambling), higher \(x_{ht}\) values yield greater energy storage for fixed \(x_{h0}\). Conversely, in the complementary region \(x_{ht} < x_{h0}\), maximal energy storage occurs when \(x_{ht}\) is minimized for given \(x_{h0}\). This asymmetric behavior stems directly from the quench-induced non-commutativity, which scales with \(|\Delta x_h|\).
    
    In fact, for a quantum battery system, the internally stored energy cannot necessarily be fully extracted via a unitary process~\cite{Nieuwenhuizen:2004exd,Alicki:2013cwy}. For a reversible unitary evolution, the maximum extractable energy from the battery is termed ergotropy, defined as
    \begin{equation}
    	\mathcal{E}(\tau) = \mathrm{Tr}[\rho(\tau) H_0] - \mathrm{Tr}[\rho_p H_0],
    \end{equation}
     where $\rho_p$ is the passive state associated with $\rho(\tau)$. Since the passive state of the final state commutes with the Hamiltonian $H_0$—implying zero energy variation. Consequently, it can be expanded in the eigenbasis $\{|i\rangle\}$ of $H_0$ as 
     \begin{equation}
     	\rho_p = \sum_i r_i |i\rangle\langle i|,
     \end{equation}
     where $H_0 = \sum_i \varepsilon_i |i\rangle\langle i|$ with $\varepsilon_{i+1} \ge \varepsilon_i$, $r_{i+1} \le r_i$, and the $r_i$ are the eigenvalues of $\rho(\tau)$~\cite{Nieuwenhuizen:2004exd}. Thus, the ergotropy is computed as
     \begin{equation}
     	\mathcal{E}(\tau) = \mathrm{Tr}[\rho(\tau) H_0] - \sum_i r_i \varepsilon_i.
     \end{equation}  
	In Fig. \ref{fig_ergo}, we show the maximum ergotropy $\mathcal{E}_\mathrm{max}$ encountered during the system's temporal evolution, normalized by the energy bandwidth $E_b$ (i.e., \(\mathcal{E} \rightarrow \mathcal{E} / E_b\) ), for various combinations of the initial and quenched scrambling parameters $x_{h0}$ and $x_{ht}$. Our aforementioned analysis holds for $\mathcal{E}_\mathrm{max}$ as well, with its behavior closely mirroring that of $E_\mathrm{max}$. This arises because the evolution involves only a perturbation to the parameter $x_h$, resulting in a minor deviation from the initial ground state and thus an energy for the passive state that closely approximates the ground-state value.
	\begin{figure}[t]
            \centering{\includegraphics[width=0.98\linewidth]{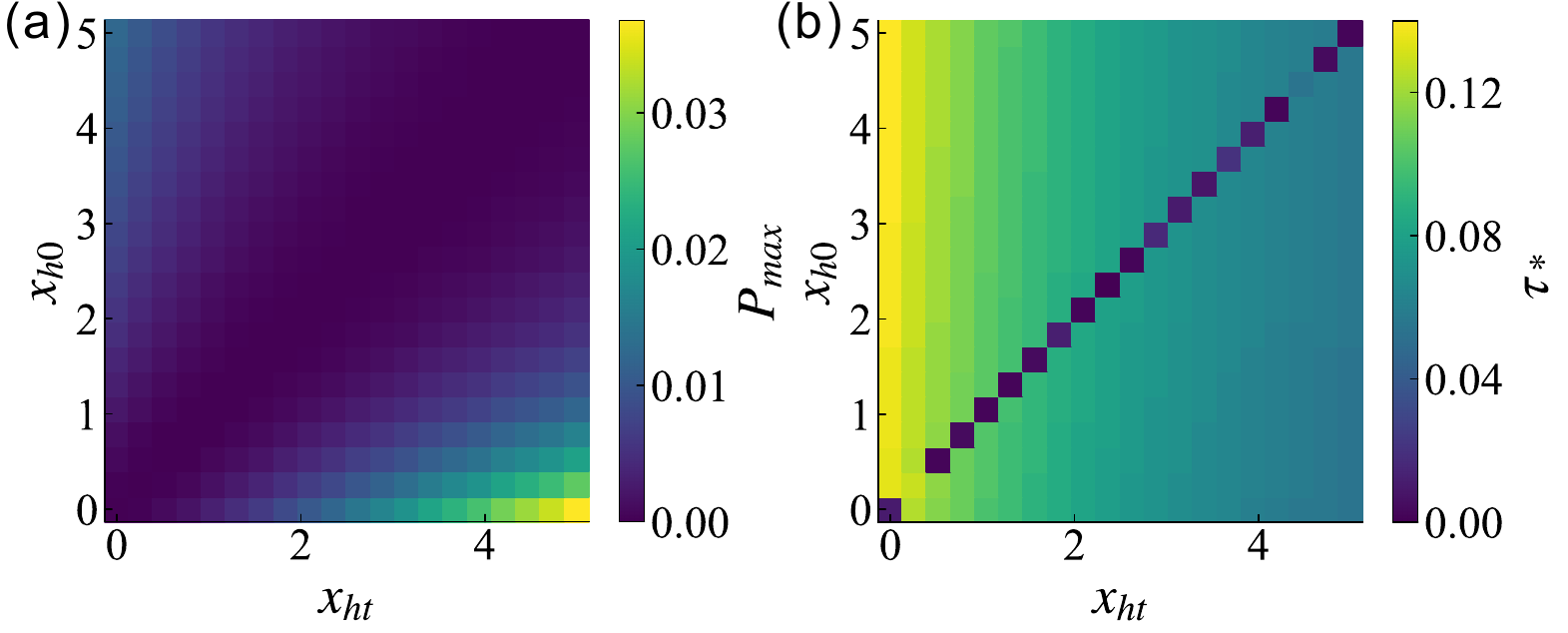}
            \caption{Impact of quenched scrambling parameters on different performance metrics: (a) maximum charging power $P_{\text{max}}$ and (b) optimal charging time $\tau_{*}$. The system size is fixed at $L=250$. }\label{fig_4}
            }
        \end{figure}

    Owing to the property of chaos rapidly scrambling system information, we are more interested in the effect of this scrambling behavior on charging efficiency. In Fig.~\ref{fig_4}, we illustrate the impact of the initial scrambling parameter and the scrambling parameter during the charging process on the maximum charging power and the optimal charging time of the system. For the numerical calculations, we set the range of both scrambling parameters to \([0,5]\). The maximum charging power, as shown in Fig.~\ref{fig_4}(a), exhibits a trend similar to that of the maximum stored energy as the parameters vary. The parameter space for the maximum power is divided into two regions by a low-value diagonal band, and the maximum power increases significantly as the difference between the two parameters grows. The maximum power is given by \(P_{\rm max} = \Delta E_{\rm max} / \tau_*\), and due to the introduction of the optimal charging time in the denominator, the value difference between the two regions across the diagonal becomes more pronounced. This results in higher maximum power being more concentrated in the lower-right region where \(x_{ht} > x_{h0}\). Specifically, when \(x_{h0}\) is smaller and \(x_{ht}\) is larger, the corresponding maximum power is significantly enhanced. 

    This behavior can be further understood through the optimal charging time, as shown in Fig.~\ref{fig_4}(b). The optimal charging time no longer exhibits symmetry across the diagonal (as the initial scrambling parameter has almost no influence on it) but instead is nearly monotonically modulated by the scrambling parameter \(x_{ht}\) during the charging process. As the scrambling intensity in the charging process increases, the optimal charging time of the system decreases progressively. Consequently, this reduction in the optimal charging time with increasing \(x_{ht}\) leads to a more significant enhancement of the maximum charging power in the direction of increasing \(x_{ht}\), which indicates a signal of accelerated charging.
    \begin{figure}[b]
            \centering{\includegraphics[width=0.98\linewidth]{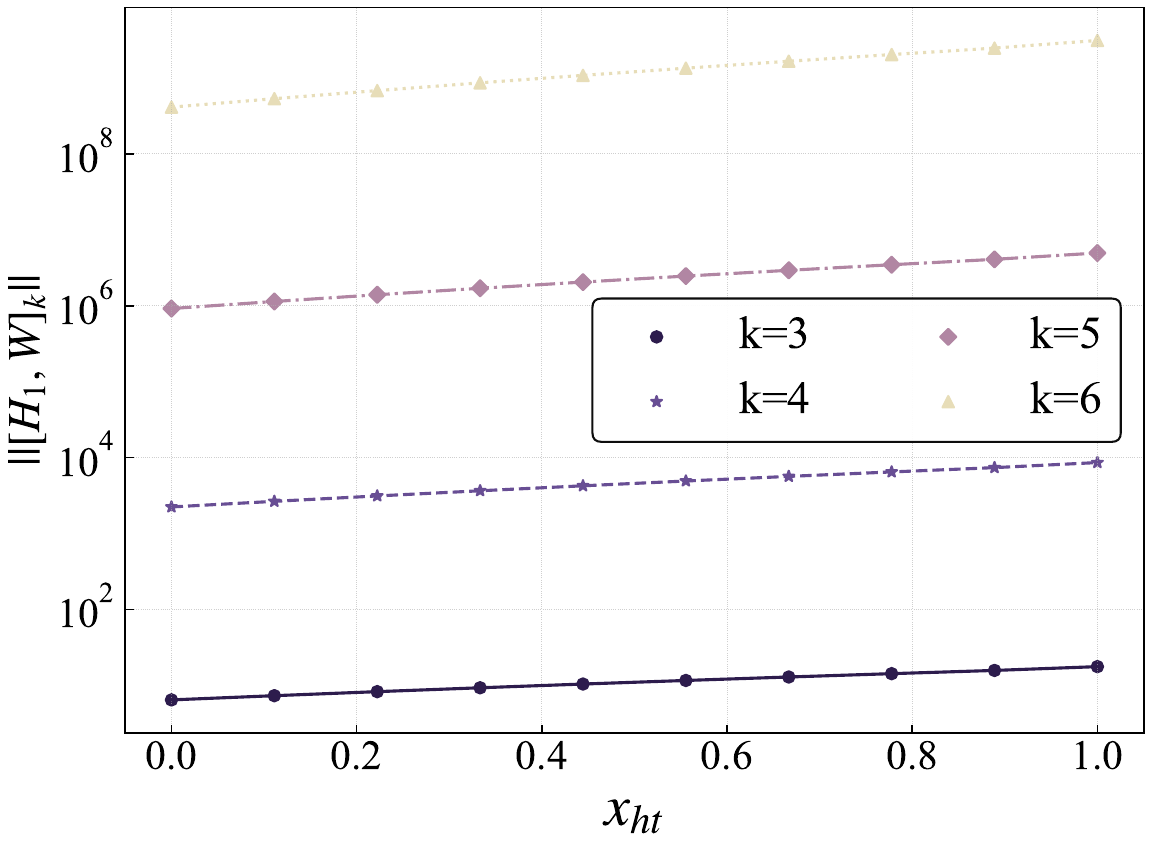}
            \caption{Operator norm growth $\|[H,W]_k\|$ versus charging scrambling parameter $x_{ht}$ for commutator orders $k=3$ (circles), $k=4$ (pentagrams), $k=5$ (diamonds), and $k=6$ (triangles). The system size is fixed at $L=250$.}
            \label{fig_5}}
        \end{figure}   

    To explain the potential acceleration of charging effects brought about by chaotic behavior, we aim to demonstrate how this chaotic behavior influences the charging process using a nested commutator similar to the butterfly effect \cite{PhysRevLett.117.091602,Swingle:2018ekw,Romero:2024wgt}. For a local operator in the Heisenberg picture, it can be expressed as: 
        \begin{equation}
             W(t) = e^{iHt} W(0) e^{-iHt}
        \end{equation}
    This can be expanded in terms of the Hamiltonian using the Baker-Campbell-Hausdorff formula as:
        \begin{equation}
              W(t) = \sum_{k}\frac{(it)^k}{k!} [H, W]_k, 
        \end{equation}
    where $[H, W]_k = [H, [H, W]_{k-1}]$. Given that the Hamiltonian \(H_{\text{sys}}\) only includes nearest-neighbor hopping interactions, the initially localized perturbation will be propagated throughout the entire system over time and will become increasingly complex. The higher-order terms in the expansion will also start to become more intricate and significant.
        
    Here, we numerically computed the maximum singular values of the various high-order nested commutators, denoted as \(\|[H, W]_k\|\), as a function of the scrambling intensity during the charging process, where the order of the nested commutator \(k \in [3, 6]\). As shown in Fig.~\ref{fig_5}, the vertical axis of the graph has been processed logarithmically. Each order of the nested commutator exhibits exponential growth with increasing scrambling intensity. Consequently, this information scrambling behavior, as the parameter \(x_{ht}\) continuously increases, leads to a reduction in the time required to achieve optimal charging.

\section{Discussion}\label{Sec_IV}
    Based on this nearest-neighbor hopping spin chain capable of simulating curved spacetime, which can effectively illustrate the chaotic behavior induced by such curvature, we investigate the impact of chaotic scrambling resulting from black hole spacetimes in an AdS\(_2\) background on the charging process of quantum batteries. In this gravitational analogue system, the scrambling behavior is solely dependent on the distribution of the hopping interactions, unlike in the SYK model, where the scrambling relies on the size of the system. This allows us to control variables to independently explore the effects of this chaotic behavior on the charging process of the quantum battery system. To ensure the validity of the scrambling parameters in the gravitational analogue system, we employ a quench protocol that compares a large number of qubit systems, which can be numerically solved using the Hubbard model.
    \begin{figure}[t]
            \centering{\includegraphics[width=0.98\linewidth]{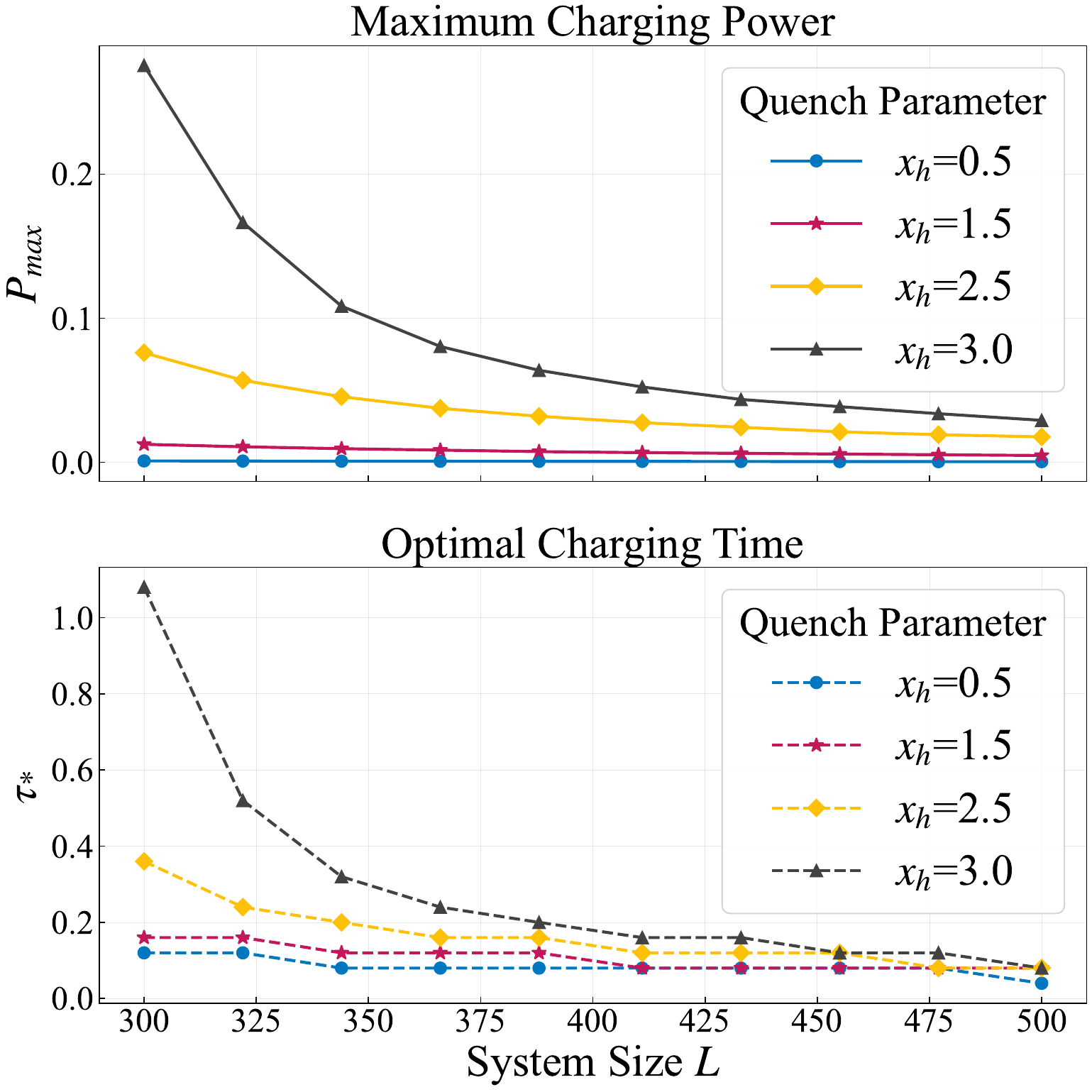}
            \caption{ Effect of system size on charging performance. The initial scrambling parameter is fixed at \( x_{h0} = 0 \), while the scrambling parameters during the quench charging process are set to \( 0.5, 1.5, 2.5, \) and \( 3.0 \), corresponding to line markers of circles, pentagons, diamonds, and triangles, respectively. The maximum charging power is represented by solid lines, and the optimal charging time is indicated by dashed lines.}\label{fig_6}
            }
        \end{figure}

    Here, we primarily discuss several issues related to system size and bandwidth regularization $H \to H/\|H\|$, where $\|\cdot\|$ denotes the norm corresponding to the maximum singular value, \cite{Rossini:2019nfu,Goold:2015cqg,PhysRevLett.118.150601}. First, to demonstrate the potential impact of system size on the charging process, we have illustrated in Fig.~\ref{fig_6} the effects of different system sizes on both the maximum charging power and the optimal charging time. In our numerical calculations, we fixed the initial scrambling parameter at \( x_{h0} = 0 \) and set the scrambling parameters during the charging process to \( 0.5 \) (circles), \( 1.5 \) (pentagons), \( 2.5 \) (diamonds), and \( 3.0 \) (triangles). The results indicate that both the maximum charging power and the optimal charging time decrease with an increase in system size. This observation is consistent with expectations, as larger qubit systems inherently possess broader energy bandwidths and larger Hilbert spaces. Consequently, under constant scrambling intensity, the charging power experiences a significant decline. However, as the system size increases, the interaction among more and larger edge qubits enables the system to reach maximum charging power more rapidly. This suggests that the influence of system size is not negligible.
     
    In the SYK quantum battery, smaller scrambling behaviors necessitate larger system sizes, while increasing system size introduces more complex interactions, leading to a combined effect of size and scrambling behavior that accelerates the optimal charging time as the scrambling decreases. Recently, Romero et al. demonstrated that the introduction of bandwidth renormalization not only accelerates the optimal charging time but also enhances the maximum charging power for SYK batteries \cite{Romero:2024wgt}. In the gravitational simulation system presented here, the scrambling behavior of the quantum battery does not depend on the system size, thus eliminating the need for complex methods of bandwidth renormalization.  
    
    In fact, if bandwidth regularization is forcibly applied to the system, further clarification is warranted. The widely proposed bandwidth regularization scheme entails dividing the charging Hamiltonian by its maximum singular value, which implies a multiplicative scaling of the Hamiltonian. This scaling directly influences the distribution of hopping interactions, resulting in a corresponding multiplicative scaling of these interactions.
    This behavior influences the scrambling parameters of the system exclusively through this scaling mechanism. Consequently, the repercussions of bandwidth regularization will lead to a significant overall reduction in the effective scrambling parameters. This diminution may impede the system's capacity to attain the anticipated maximum charging power within the effective scrambling time and under the constraints imposed by the boundary effects arising from finite size, thereby causing the optimal charging time to become contingent upon the total charging period. If the charging time is extended further, the scrambling dynamics of the system may either lose their effectiveness or the boundary effects related to finite size may undermine the efficacy of the analogue system. Nevertheless, we can still investigate the influence of the scrambling behavior during the effective time on the maximum charging power.
        \begin{figure}[t]
            \centering{\includegraphics[width=0.98\linewidth]{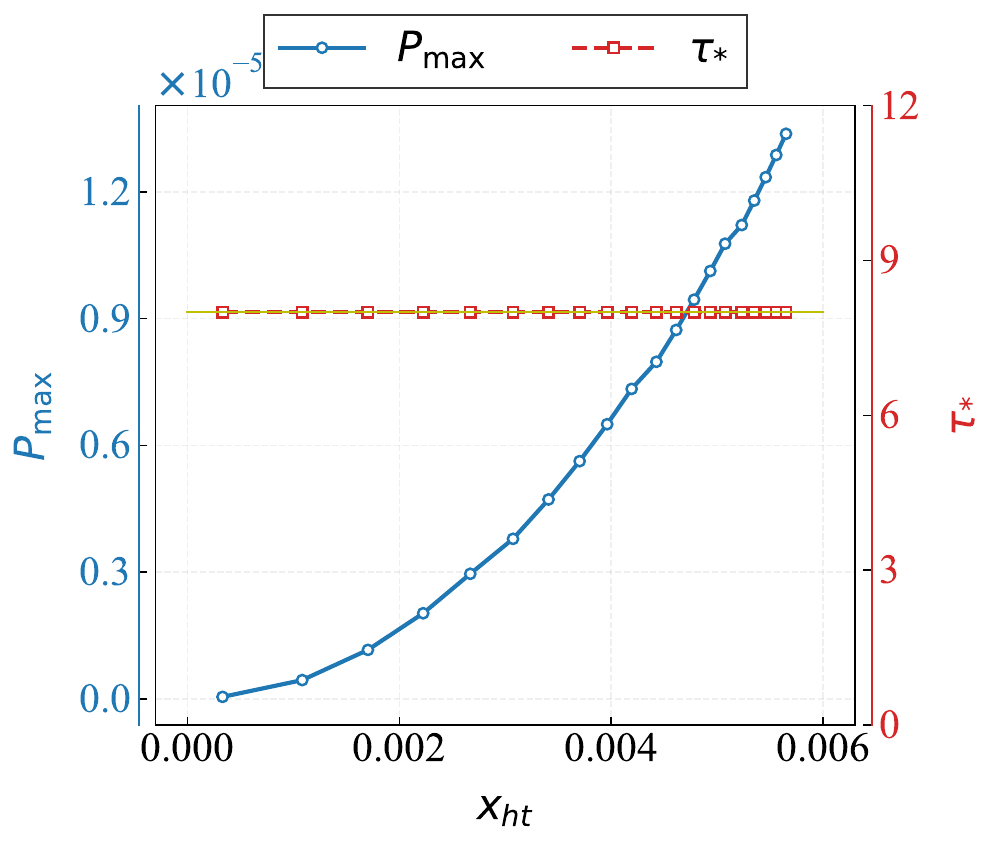}
            \caption{ The impact of bandwidth regularization on the charging process. We set \(x_{h0} = 0\), with the maximum charging power indicated by the blue circle and the optimal charging time represented by the red square. The system size is fixed at \(L = 250\). }\label{fig_7}
            }
        \end{figure}

    In Fig.\ref{fig_7}, we present the numerical calculations of the maximum charging power (blue circles) and the optimal charging time (red squares) following the bandwidth regularization of the charging Hamiltonian, plotted against the corresponding scrambling parameters post-regularization. The yellow solid line denotes the total charging time. We set the initial scrambling parameter \(x_{h0} = 0\). The results indicate that the optimal charging time remains consistent with the total charging time (yellow solid line), primarily due to the relatively small scrambling parameters, which fail to achieve the anticipated maximum charging power within the effective scrambling time. Additionally, we observe that the maximum charging power increases with the scrambling parameter, indicating that the bandwidth regularization does not alter the scrambling effects within the system.

\section{Conclusion}\label{Sec_V}
    Research on quantum batteries has focused intensely on enhancing key performance metrics—including energy storage capacity, charging rate, and operational stability—as these advancements critically determine the feasibility of deploying such theoretical constructs in future energy technologies. Quantum information scrambling, characterized by rapid propagation of local perturbations throughout a system, exhibits significant potential for improving quantum battery charging dynamics. Motivated by recent insights from AdS/CFT correspondence establishing black holes as maximal quantum information scramblers, and leveraging experimental demonstrations of curved spacetime quantum effects simulated via site-dependent isotropic XY models on superconducting quantum processors, we investigate scrambling dynamics in quantum battery charging through controlled scrambling parameter quenches within curved spacetime-analog spin chains. This work specifically examines how black hole spacetime-induced scrambling affects three critical performance benchmarks: maximum stored energy $E_{\max}$, maximum charging power $P_{\max}$, and optimal charging time $\tau_{*}$.
   
    Our numerical results demonstrate that the quantum battery system exhibits enhanced maximum stored energy and maximum charging power with increasing quench-induced scrambling parameter variation $\Delta x_h \equiv x_{ht} - x_{h0}$. Notably, the performance enhancement becomes increasingly pronounced in the parameter regime where $x_{ht} > x_{h0}$. For the optimal charging time $\tau_{*}$, we observe insensitivity to the initial scrambling parameter $x_{h0}$, while it displays significant sensitivity to the charging parameter $x_{ht}$, exhibiting monotonic decay as $x_{ht}$ increases. This behavior may indicate accelerated charging dynamics.
    These phenomena are interpretable through the lens of quantum chaos. Specifically, the exponential growth of OTOCs, a hallmark of quantum scrambling, correlates positively with enhanced energy transfer rates within this quantum battery architecture. This suggests that stronger scrambling behavior accelerates the system's transition from its initial state to higher energy eigenstates.
    It is noteworthy that the bandwidth regularization from \( H \) $\to $\( H / \| H \| \) significantly diminishes the effective scrambling intensity of the system. However, this process does not alter the trend whereby the maximum charging power of the quantum battery increases with the growing scrambling intensity of the system.
    
    Our study demonstrates that black hole-inspired chaotic systems provide a versatile platform for optimizing quantum battery performance. By engineering the scrambling dynamics through tailored hopping interactions distributions, we can control key scrambling parameters to thoroughly investigate battery performance metrics such as energy storage capacity, charging power, and optimal charging times. These insights not only enhance our understanding of the interplay between quantum information scrambling and energy transfer but also pave the way for designing advanced quantum energy storage devices. 
    Moreover, these simulation-based quantum batteries can offer new perspectives for research in this field, linking specific behaviors across different theories and fostering the integration and profound understanding of various disciplines.

\acknowledgments
    This work was supported by the National Natural Science Foundation of China under Grants No. 12475051,  No. 12035005,  and No. 12421005; the science and technology innovation Program of Hunan Province under grant No. 2024RC1050; the Natural Science Foundation of Hunan Province under grant No. 2023JJ30384; and the innovative research group of Hunan Province under Grant No. 2024JJ1006. ZT was supported by the National Natural Science Foundation of China under Grant No. 12575050, and the scientific research start-up funds of Hangzhou Normal University: 4245C50224204016.



\begin{thebibliography}{92}%
	\makeatletter
	\providecommand \@ifxundefined [1]{%
		\@ifx{#1\undefined}
	}%
	\providecommand \@ifnum [1]{%
		\ifnum #1\expandafter \@firstoftwo
		\else \expandafter \@secondoftwo
		\fi
	}%
	\providecommand \@ifx [1]{%
		\ifx #1\expandafter \@firstoftwo
		\else \expandafter \@secondoftwo
		\fi
	}%
	\providecommand \natexlab [1]{#1}%
	\providecommand \enquote  [1]{``#1''}%
	\providecommand \bibnamefont  [1]{#1}%
	\providecommand \bibfnamefont [1]{#1}%
	\providecommand \citenamefont [1]{#1}%
	\providecommand \href@noop [0]{\@secondoftwo}%
	\providecommand \href [0]{\begingroup \@sanitize@url \@href}%
	\providecommand \@href[1]{\@@startlink{#1}\@@href}%
	\providecommand \@@href[1]{\endgroup#1\@@endlink}%
	\providecommand \@sanitize@url [0]{\catcode `\\12\catcode `\$12\catcode
		`\&12\catcode `\#12\catcode `\^12\catcode `\_12\catcode `\%12\relax}%
	\providecommand \@@startlink[1]{}%
	\providecommand \@@endlink[0]{}%
	\providecommand \url  [0]{\begingroup\@sanitize@url \@url }%
	\providecommand \@url [1]{\endgroup\@href {#1}{\urlprefix }}%
	\providecommand \urlprefix  [0]{URL }%
	\providecommand \Eprint [0]{\href }%
	\providecommand \doibase [0]{https://doi.org/}%
	\providecommand \selectlanguage [0]{\@gobble}%
	\providecommand \bibinfo  [0]{\@secondoftwo}%
	\providecommand \bibfield  [0]{\@secondoftwo}%
	\providecommand \translation [1]{[#1]}%
	\providecommand \BibitemOpen [0]{}%
	\providecommand \bibitemStop [0]{}%
	\providecommand \bibitemNoStop [0]{.\EOS\space}%
	\providecommand \EOS [0]{\spacefactor3000\relax}%
	\providecommand \BibitemShut  [1]{\csname bibitem#1\endcsname}%
	\let\auto@bib@innerbib\@empty
	\bibitem [{\citenamefont {Pirandola}\ \emph {et~al.}(2017)\citenamefont
		{Pirandola}, \citenamefont {Laurenza}, \citenamefont {Ottaviani},\ and\
		\citenamefont {Banchi}}]{Pirandola:2017kzd}%
	\BibitemOpen
	\bibfield  {author} {\bibinfo {author} {\bibfnamefont {S.}~\bibnamefont
			{Pirandola}}, \bibinfo {author} {\bibfnamefont {R.}~\bibnamefont {Laurenza}},
		\bibinfo {author} {\bibfnamefont {C.}~\bibnamefont {Ottaviani}},\ and\
		\bibinfo {author} {\bibfnamefont {L.}~\bibnamefont {Banchi}},\ }\bibfield
	{title} {\bibinfo {title} {{Fundamental limits of repeaterless quantum
				communications}},\ }\href {https://doi.org/10.1038/ncomms15043} {\bibfield
		{journal} {\bibinfo  {journal} {Nature Commun.}\ }\textbf {\bibinfo {volume}
			{8}},\ \bibinfo {pages} {15043} (\bibinfo {year} {2017})}\BibitemShut
	{NoStop}%
	\bibitem [{\citenamefont {Liu}\ \emph {et~al.}(2022)\citenamefont {Liu},
		\citenamefont {Wen}, \citenamefont {Tian}, \citenamefont {Jing},\ and\
		\citenamefont {Wang}}]{Liu:2021frj}%
	\BibitemOpen
	\bibfield  {author} {\bibinfo {author} {\bibfnamefont {Q.}~\bibnamefont
			{Liu}}, \bibinfo {author} {\bibfnamefont {C.}~\bibnamefont {Wen}}, \bibinfo
		{author} {\bibfnamefont {Z.}~\bibnamefont {Tian}}, \bibinfo {author}
		{\bibfnamefont {J.}~\bibnamefont {Jing}},\ and\ \bibinfo {author}
		{\bibfnamefont {J.}~\bibnamefont {Wang}},\ }\bibfield  {title} {\bibinfo
		{title} {{Gravity-enhanced quantum spatial target detection}},\ }\href
	{https://doi.org/10.1103/PhysRevA.105.062428} {\bibfield  {journal} {\bibinfo
			{journal} {Phys. Rev. A}\ }\textbf {\bibinfo {volume} {105}},\ \bibinfo
		{pages} {062428} (\bibinfo {year} {2022})},\ \Eprint
	{https://arxiv.org/abs/2104.02314} {arXiv:2104.02314 [gr-qc]} \BibitemShut
	{NoStop}%
	\bibitem [{\citenamefont {Zhang}\ and\ \citenamefont
		{Zhuang}(2021)}]{Zhang:2020skj}%
	\BibitemOpen
	\bibfield  {author} {\bibinfo {author} {\bibfnamefont {Z.}~\bibnamefont
			{Zhang}}\ and\ \bibinfo {author} {\bibfnamefont {Q.}~\bibnamefont {Zhuang}},\
	}\bibfield  {title} {\bibinfo {title} {{Distributed quantum sensing}},\
	}\href {https://doi.org/10.1088/2058-9565/abd4c3} {\bibfield  {journal}
		{\bibinfo  {journal} {Quantum Sci. Technol.}\ }\textbf {\bibinfo {volume}
			{6}},\ \bibinfo {pages} {043001} (\bibinfo {year} {2021})},\ \Eprint
	{https://arxiv.org/abs/2010.14744} {arXiv:2010.14744 [quant-ph]} \BibitemShut
	{NoStop}%
	\bibitem [{\citenamefont {Degen}\ \emph {et~al.}(2017)\citenamefont {Degen},
		\citenamefont {Reinhard},\ and\ \citenamefont {Cappellaro}}]{Degen:2016pxo}%
	\BibitemOpen
	\bibfield  {author} {\bibinfo {author} {\bibfnamefont {C.~L.}\ \bibnamefont
			{Degen}}, \bibinfo {author} {\bibfnamefont {F.}~\bibnamefont {Reinhard}},\
		and\ \bibinfo {author} {\bibfnamefont {P.}~\bibnamefont {Cappellaro}},\
	}\bibfield  {title} {\bibinfo {title} {{Quantum sensing}},\ }\href
	{https://doi.org/10.1103/RevModPhys.89.035002} {\bibfield  {journal}
		{\bibinfo  {journal} {Rev. Mod. Phys.}\ }\textbf {\bibinfo {volume} {89}},\
		\bibinfo {pages} {035002} (\bibinfo {year} {2017})},\ \Eprint
	{https://arxiv.org/abs/1611.02427} {arXiv:1611.02427 [quant-ph]} \BibitemShut
	{NoStop}%
	\bibitem [{\citenamefont {Wang}\ \emph {et~al.}(2019)\citenamefont {Wang},
		\citenamefont {Liu}, \citenamefont {Jing},\ and\ \citenamefont
		{Chen}}]{Wang:2019aqr}%
	\BibitemOpen
	\bibfield  {author} {\bibinfo {author} {\bibfnamefont {J.}~\bibnamefont
			{Wang}}, \bibinfo {author} {\bibfnamefont {T.}~\bibnamefont {Liu}}, \bibinfo
		{author} {\bibfnamefont {J.}~\bibnamefont {Jing}},\ and\ \bibinfo {author}
		{\bibfnamefont {S.}~\bibnamefont {Chen}},\ }\bibfield  {title} {\bibinfo
		{title} {{Synchronization and estimation of gravity-induced time difference
				for quantum clocks}},\ }\href {https://doi.org/10.1002/qute.201900003}
	{\bibfield  {journal} {\bibinfo  {journal} {Adv. Quantum Technol.}\ }\textbf
		{\bibinfo {volume} {2}},\ \bibinfo {pages} {1900003} (\bibinfo {year}
		{2019})},\ \Eprint {https://arxiv.org/abs/1911.05279} {arXiv:1911.05279
		[quant-ph]} \BibitemShut {NoStop}%
	\bibitem [{\citenamefont {Zhang}\ \emph {et~al.}(2019)\citenamefont {Zhang},
		\citenamefont {Jing}, \citenamefont {Fan},\ and\ \citenamefont
		{Wang}}]{Zhang:2018atw}%
	\BibitemOpen
	\bibfield  {author} {\bibinfo {author} {\bibfnamefont {L.}~\bibnamefont
			{Zhang}}, \bibinfo {author} {\bibfnamefont {J.}~\bibnamefont {Jing}},
		\bibinfo {author} {\bibfnamefont {H.}~\bibnamefont {Fan}},\ and\ \bibinfo
		{author} {\bibfnamefont {J.}~\bibnamefont {Wang}},\ }\bibfield  {title}
	{\bibinfo {title} {{Multipartite Quantum Clock Synchronization under The
				Influence of Unruh Thermal Noise}},\ }\href
	{https://doi.org/10.1002/andp.201900067} {\bibfield  {journal} {\bibinfo
			{journal} {Annalen Phys.}\ }\textbf {\bibinfo {volume} {531}},\ \bibinfo
		{pages} {1900067} (\bibinfo {year} {2019})},\ \Eprint
	{https://arxiv.org/abs/1801.01666} {arXiv:1801.01666 [quant-ph]} \BibitemShut
	{NoStop}%
	\bibitem [{\citenamefont {Wang}\ \emph {et~al.}(2016)\citenamefont {Wang},
		\citenamefont {Tian}, \citenamefont {Jing},\ and\ \citenamefont
		{Fan}}]{Wang:2015yma}%
	\BibitemOpen
	\bibfield  {author} {\bibinfo {author} {\bibfnamefont {J.}~\bibnamefont
			{Wang}}, \bibinfo {author} {\bibfnamefont {Z.}~\bibnamefont {Tian}}, \bibinfo
		{author} {\bibfnamefont {J.}~\bibnamefont {Jing}},\ and\ \bibinfo {author}
		{\bibfnamefont {H.}~\bibnamefont {Fan}},\ }\bibfield  {title} {\bibinfo
		{title} {{Influence of relativistic effects on satellite-based clock
				synchronization}},\ }\href {https://doi.org/10.1103/PhysRevD.93.065008}
	{\bibfield  {journal} {\bibinfo  {journal} {Phys. Rev. D}\ }\textbf {\bibinfo
			{volume} {93}},\ \bibinfo {pages} {065008} (\bibinfo {year} {2016})},\
	\Eprint {https://arxiv.org/abs/1501.01478} {arXiv:1501.01478 [quant-ph]}
	\BibitemShut {NoStop}%
	\bibitem [{\citenamefont {Garms}\ \emph {et~al.}(2024)\citenamefont {Garms},
		\citenamefont {Para\"\i{}so}, \citenamefont {Hanley}, \citenamefont {Khalid},
		\citenamefont {Rafferty}, \citenamefont {Grant}, \citenamefont {Newman},
		\citenamefont {Shields}, \citenamefont {Cid},\ and\ \citenamefont
		{O'Neill}}]{Garms:2024hip}%
	\BibitemOpen
	\bibfield  {author} {\bibinfo {author} {\bibfnamefont {L.}~\bibnamefont
			{Garms}}, \bibinfo {author} {\bibfnamefont {T.~K.}\ \bibnamefont
			{Para\"\i{}so}}, \bibinfo {author} {\bibfnamefont {N.}~\bibnamefont
			{Hanley}}, \bibinfo {author} {\bibfnamefont {A.}~\bibnamefont {Khalid}},
		\bibinfo {author} {\bibfnamefont {C.}~\bibnamefont {Rafferty}}, \bibinfo
		{author} {\bibfnamefont {J.}~\bibnamefont {Grant}}, \bibinfo {author}
		{\bibfnamefont {J.}~\bibnamefont {Newman}}, \bibinfo {author} {\bibfnamefont
			{A.~J.}\ \bibnamefont {Shields}}, \bibinfo {author} {\bibfnamefont
			{C.}~\bibnamefont {Cid}},\ and\ \bibinfo {author} {\bibfnamefont
			{M.}~\bibnamefont {O'Neill}},\ }\bibfield  {title} {\bibinfo {title}
		{{Experimental Integration of Quantum Key Distribution and Post-Quantum
				Cryptography in a Hybrid Quantum-Safe Cryptosystem}},\ }\href
	{https://doi.org/10.1002/qute.202300304} {\bibfield  {journal} {\bibinfo
			{journal} {Adv. Quantum Technol.}\ }\textbf {\bibinfo {volume} {7}},\
		\bibinfo {pages} {2300304} (\bibinfo {year} {2024})}\BibitemShut {NoStop}%
	\bibitem [{\citenamefont {Joseph}\ \emph {et~al.}(2022)\citenamefont {Joseph},
		\citenamefont {Misoczki}, \citenamefont {Manzano}, \citenamefont {Tricot},
		\citenamefont {Pinuaga}, \citenamefont {Lacombe}, \citenamefont
		{Leichenauer}, \citenamefont {Hidary}, \citenamefont {Venables},\ and\
		\citenamefont {Hansen}}]{Joseph:2022kac}%
	\BibitemOpen
	\bibfield  {author} {\bibinfo {author} {\bibfnamefont {D.}~\bibnamefont
			{Joseph}}, \bibinfo {author} {\bibfnamefont {R.}~\bibnamefont {Misoczki}},
		\bibinfo {author} {\bibfnamefont {M.}~\bibnamefont {Manzano}}, \bibinfo
		{author} {\bibfnamefont {J.}~\bibnamefont {Tricot}}, \bibinfo {author}
		{\bibfnamefont {F.~D.}\ \bibnamefont {Pinuaga}}, \bibinfo {author}
		{\bibfnamefont {O.}~\bibnamefont {Lacombe}}, \bibinfo {author} {\bibfnamefont
			{S.}~\bibnamefont {Leichenauer}}, \bibinfo {author} {\bibfnamefont
			{J.}~\bibnamefont {Hidary}}, \bibinfo {author} {\bibfnamefont
			{P.}~\bibnamefont {Venables}},\ and\ \bibinfo {author} {\bibfnamefont
			{R.}~\bibnamefont {Hansen}},\ }\bibfield  {title} {\bibinfo {title}
		{{Transitioning organizations to post-quantum cryptography}},\ }\href
	{https://doi.org/10.1038/s41586-022-04623-2} {\bibfield  {journal} {\bibinfo
			{journal} {Nature}\ }\textbf {\bibinfo {volume} {605}},\ \bibinfo {pages}
		{237} (\bibinfo {year} {2022})}\BibitemShut {NoStop}%
	\bibitem [{\citenamefont {Portmann}\ and\ \citenamefont
		{Renner}(2022)}]{Portmann:2021byc}%
	\BibitemOpen
	\bibfield  {author} {\bibinfo {author} {\bibfnamefont {C.}~\bibnamefont
			{Portmann}}\ and\ \bibinfo {author} {\bibfnamefont {R.}~\bibnamefont
			{Renner}},\ }\bibfield  {title} {\bibinfo {title} {{Security in quantum
				cryptography}},\ }\href {https://doi.org/10.1103/RevModPhys.94.025008}
	{\bibfield  {journal} {\bibinfo  {journal} {Rev. Mod. Phys.}\ }\textbf
		{\bibinfo {volume} {94}},\ \bibinfo {pages} {025008} (\bibinfo {year}
		{2022})},\ \Eprint {https://arxiv.org/abs/2102.00021} {arXiv:2102.00021
		[quant-ph]} \BibitemShut {NoStop}%
	\bibitem [{\citenamefont {Aziz}\ and\ \citenamefont
		{Howl}(2025)}]{Aziz:2025ypo}%
	\BibitemOpen
	\bibfield  {author} {\bibinfo {author} {\bibfnamefont {J.}~\bibnamefont
			{Aziz}}\ and\ \bibinfo {author} {\bibfnamefont {R.}~\bibnamefont {Howl}},\
	}\bibfield  {title} {\bibinfo {title} {{Classical theories of gravity produce
				entanglement}},\ }\href {https://doi.org/10.1038/s41586-025-09595-7}
	{\bibfield  {journal} {\bibinfo  {journal} {Nature}\ }\textbf {\bibinfo
			{volume} {646}},\ \bibinfo {pages} {813} (\bibinfo {year} {2025})},\ \Eprint
	{https://arxiv.org/abs/2510.19714} {arXiv:2510.19714 [quant-ph]} \BibitemShut
	{NoStop}%
	\bibitem [{\citenamefont {Marletto}\ and\ \citenamefont
		{Vedral}(2025)}]{Marletto:2024ltk}%
	\BibitemOpen
	\bibfield  {author} {\bibinfo {author} {\bibfnamefont {C.}~\bibnamefont
			{Marletto}}\ and\ \bibinfo {author} {\bibfnamefont {V.}~\bibnamefont
			{Vedral}},\ }\bibfield  {title} {\bibinfo {title} {{Quantum-information
				methods for quantum gravity laboratory-based tests}},\ }\href
	{https://doi.org/10.1103/RevModPhys.97.015006} {\bibfield  {journal}
		{\bibinfo  {journal} {Rev. Mod. Phys.}\ }\textbf {\bibinfo {volume} {97}},\
		\bibinfo {pages} {015006} (\bibinfo {year} {2025})},\ \Eprint
	{https://arxiv.org/abs/2410.07262} {arXiv:2410.07262 [quant-ph]} \BibitemShut
	{NoStop}%
	\bibitem [{\citenamefont {He}\ \emph {et~al.}(2025)\citenamefont {He},
		\citenamefont {Huang}, \citenamefont {Zhang}, \citenamefont {Liu},
		\citenamefont {Zhang}, \citenamefont {Feng}, \citenamefont {Liu},
		\citenamefont {Cui}, \citenamefont {Huang}, \citenamefont {Zhang},\ and\
		\citenamefont {Zhang}}]{article}%
	\BibitemOpen
	\bibfield  {author} {\bibinfo {author} {\bibfnamefont {L.}~\bibnamefont
			{He}}, \bibinfo {author} {\bibfnamefont {L.}~\bibnamefont {Huang}}, \bibinfo
		{author} {\bibfnamefont {W.}~\bibnamefont {Zhang}}, \bibinfo {author}
		{\bibfnamefont {D.}~\bibnamefont {Liu}}, \bibinfo {author} {\bibfnamefont
			{H.}~\bibnamefont {Zhang}}, \bibinfo {author} {\bibfnamefont
			{X.}~\bibnamefont {Feng}}, \bibinfo {author} {\bibfnamefont {F.}~\bibnamefont
			{Liu}}, \bibinfo {author} {\bibfnamefont {K.}~\bibnamefont {Cui}}, \bibinfo
		{author} {\bibfnamefont {Y.}~\bibnamefont {Huang}}, \bibinfo {author}
		{\bibfnamefont {W.}~\bibnamefont {Zhang}},\ and\ \bibinfo {author}
		{\bibfnamefont {X.}~\bibnamefont {Zhang}},\ }\bibfield  {title} {\bibinfo
		{title} {Hyperbolic topological quantum sources},\ }\href
	{https://doi.org/10.1002/advs.202417708} {\bibfield  {journal} {\bibinfo
			{journal} {Adv. Sci.}\ }\textbf {\bibinfo {volume} {12}} (\bibinfo {year}
		{2025})}\BibitemShut {NoStop}%
	\bibitem [{\citenamefont {Feng}\ \emph {et~al.}(2025)\citenamefont {Feng},
		\citenamefont {Zhuang}, \citenamefont {Liu}, \citenamefont {Liu},
		\citenamefont {Li},\ and\ \citenamefont {Zhang}}]{Feng:2025djf}%
	\BibitemOpen
	\bibfield  {author} {\bibinfo {author} {\bibfnamefont {Z.}~\bibnamefont
			{Feng}}, \bibinfo {author} {\bibfnamefont {R.}~\bibnamefont {Zhuang}},
		\bibinfo {author} {\bibfnamefont {S.}~\bibnamefont {Liu}}, \bibinfo {author}
		{\bibfnamefont {G.}~\bibnamefont {Liu}}, \bibinfo {author} {\bibfnamefont
			{K.}~\bibnamefont {Li}},\ and\ \bibinfo {author} {\bibfnamefont
			{Y.}~\bibnamefont {Zhang}},\ }\bibfield  {title} {\bibinfo {title}
		{{Observation of Optical Precursor in Time-Energy-Entangled W Triphotons}},\
	}\href {https://doi.org/10.1002/advs.202501626} {\bibfield  {journal}
		{\bibinfo  {journal} {Adv. Sci.}\ }\textbf {\bibinfo {volume} {12}},\
		\bibinfo {pages} {2501626} (\bibinfo {year} {2025})}\BibitemShut {NoStop}%
	\bibitem [{\citenamefont {Kieu}(2004)}]{Kieu:2004fbx}%
	\BibitemOpen
	\bibfield  {author} {\bibinfo {author} {\bibfnamefont {T.~D.}\ \bibnamefont
			{Kieu}},\ }\bibfield  {title} {\bibinfo {title} {{The Second Law, Maxwell's
				Demon, and Work Derivable from Quantum Heat Engines}},\ }\href
	{https://doi.org/10.1103/PhysRevLett.93.140403} {\bibfield  {journal}
		{\bibinfo  {journal} {Phys. Rev. Lett.}\ }\textbf {\bibinfo {volume} {93}},\
		\bibinfo {pages} {140403} (\bibinfo {year} {2004})}\BibitemShut {NoStop}%
	\bibitem [{\citenamefont {Hovhannisyan}\ \emph {et~al.}(2013)\citenamefont
		{Hovhannisyan}, \citenamefont {Perarnau-Llobet}, \citenamefont {Huber},\ and\
		\citenamefont {Ac{\'\i}n}}]{Hovhannisyan:2013pqk}%
	\BibitemOpen
	\bibfield  {author} {\bibinfo {author} {\bibfnamefont {K.~V.}\ \bibnamefont
			{Hovhannisyan}}, \bibinfo {author} {\bibfnamefont {M.}~\bibnamefont
			{Perarnau-Llobet}}, \bibinfo {author} {\bibfnamefont {M.}~\bibnamefont
			{Huber}},\ and\ \bibinfo {author} {\bibfnamefont {A.}~\bibnamefont
			{Ac{\'\i}n}},\ }\bibfield  {title} {\bibinfo {title} {{Entanglement
				Generation is Not Necessary for Optimal Work Extraction}},\ }\href
	{https://doi.org/10.1103/PhysRevLett.111.240401} {\bibfield  {journal}
		{\bibinfo  {journal} {Phys. Rev. Lett.}\ }\textbf {\bibinfo {volume} {111}},\
		\bibinfo {pages} {240401} (\bibinfo {year} {2013})}\BibitemShut {NoStop}%
	\bibitem [{\citenamefont {Uzdin}\ \emph {et~al.}(2016)\citenamefont {Uzdin},
		\citenamefont {Levy},\ and\ \citenamefont {Kosloff}}]{e18040124}%
	\BibitemOpen
	\bibfield  {author} {\bibinfo {author} {\bibfnamefont {R.}~\bibnamefont
			{Uzdin}}, \bibinfo {author} {\bibfnamefont {A.}~\bibnamefont {Levy}},\ and\
		\bibinfo {author} {\bibfnamefont {R.}~\bibnamefont {Kosloff}},\ }\bibfield
	{title} {\bibinfo {title} {Quantum heat machines equivalence, work extraction
			beyond markovianity, and strong coupling via heat exchangers},\ }\bibfield
	{journal} {\bibinfo  {journal} {Entropy}\ }\textbf {\bibinfo {volume} {18}},\
	\href {https://doi.org/10.3390/e18040124} {10.3390/e18040124} (\bibinfo
	{year} {2016})\BibitemShut {NoStop}%
	\bibitem [{\citenamefont {Uzdin}\ \emph {et~al.}(2015)\citenamefont {Uzdin},
		\citenamefont {Levy},\ and\ \citenamefont {Kosloff}}]{Uzdin:2015gcs}%
	\BibitemOpen
	\bibfield  {author} {\bibinfo {author} {\bibfnamefont {R.}~\bibnamefont
			{Uzdin}}, \bibinfo {author} {\bibfnamefont {A.}~\bibnamefont {Levy}},\ and\
		\bibinfo {author} {\bibfnamefont {R.}~\bibnamefont {Kosloff}},\ }\bibfield
	{title} {\bibinfo {title} {{Equivalence of Quantum Heat Machines, and
				Quantum-Thermodynamic Signatures}},\ }\href
	{https://doi.org/10.1103/PhysRevX.5.031044} {\bibfield  {journal} {\bibinfo
			{journal} {Phys. Rev. X}\ }\textbf {\bibinfo {volume} {5}},\ \bibinfo {pages}
		{031044} (\bibinfo {year} {2015})}\BibitemShut {NoStop}%
	\bibitem [{\citenamefont {Alicki}\ and\ \citenamefont
		{Fannes}(2013)}]{Alicki:2013cwy}%
	\BibitemOpen
	\bibfield  {author} {\bibinfo {author} {\bibfnamefont {R.}~\bibnamefont
			{Alicki}}\ and\ \bibinfo {author} {\bibfnamefont {M.}~\bibnamefont
			{Fannes}},\ }\bibfield  {title} {\bibinfo {title} {{Entanglement boost for
				extractable work from ensembles of quantum batteries}},\ }\href
	{https://doi.org/10.1103/PhysRevE.87.042123} {\bibfield  {journal} {\bibinfo
			{journal} {Phys. Rev. E}\ }\textbf {\bibinfo {volume} {87}},\ \bibinfo
		{pages} {042123} (\bibinfo {year} {2013})}\BibitemShut {NoStop}%
	\bibitem [{\citenamefont {Andolina}\ \emph {et~al.}(2025)\citenamefont
		{Andolina}, \citenamefont {Stanzione}, \citenamefont {Giovannetti},\ and\
		\citenamefont {Polini}}]{Andolina:2024iet}%
	\BibitemOpen
	\bibfield  {author} {\bibinfo {author} {\bibfnamefont {G.~M.}\ \bibnamefont
			{Andolina}}, \bibinfo {author} {\bibfnamefont {V.}~\bibnamefont {Stanzione}},
		\bibinfo {author} {\bibfnamefont {V.}~\bibnamefont {Giovannetti}},\ and\
		\bibinfo {author} {\bibfnamefont {M.}~\bibnamefont {Polini}},\ }\bibfield
	{title} {\bibinfo {title} {{Genuine Quantum Advantage in Anharmonic Bosonic
				Quantum Batteries}},\ }\href {https://doi.org/10.1103/kzvn-dj7v} {\bibfield
		{journal} {\bibinfo  {journal} {Phys. Rev. Lett.}\ }\textbf {\bibinfo
			{volume} {134}},\ \bibinfo {pages} {240403} (\bibinfo {year} {2025})},\
	\Eprint {https://arxiv.org/abs/2409.08627} {arXiv:2409.08627 [quant-ph]}
	\BibitemShut {NoStop}%
	\bibitem [{\citenamefont {Hu}\ \emph {et~al.}(2025{\natexlab{a}})\citenamefont
		{Hu}, \citenamefont {Gao},\ and\ \citenamefont {Fan}}]{Hu:2025sia}%
	\BibitemOpen
	\bibfield  {author} {\bibinfo {author} {\bibfnamefont {M.-L.}\ \bibnamefont
			{Hu}}, \bibinfo {author} {\bibfnamefont {T.}~\bibnamefont {Gao}},\ and\
		\bibinfo {author} {\bibfnamefont {H.}~\bibnamefont {Fan}},\ }\bibfield
	{title} {\bibinfo {title} {Enhancing the charging performance of an atomic
			quantum battery},\ }\href
	{https://doi.org/https://doi.org/10.1002/qute.202500422} {\bibfield
		{journal} {\bibinfo  {journal} {Adv. Quantum Technol.}\ }\textbf {\bibinfo
			{volume} {8}},\ \bibinfo {pages} {e00422} (\bibinfo {year}
		{2025}{\natexlab{a}})}\BibitemShut {NoStop}%
	\bibitem [{\citenamefont {Goold}\ \emph {et~al.}(2015)\citenamefont {Goold},
		\citenamefont {Modi}, \citenamefont {Vinjanampathy},\ and\ \citenamefont
		{Binder}}]{Goold:2015cqg}%
	\BibitemOpen
	\bibfield  {author} {\bibinfo {author} {\bibfnamefont {J.}~\bibnamefont
			{Goold}}, \bibinfo {author} {\bibfnamefont {K.}~\bibnamefont {Modi}},
		\bibinfo {author} {\bibfnamefont {S.}~\bibnamefont {Vinjanampathy}},\ and\
		\bibinfo {author} {\bibfnamefont {F.~C.}\ \bibnamefont {Binder}},\ }\bibfield
	{title} {\bibinfo {title} {{Quantacell: powerful charging of quantum
				batteries}},\ }\href {https://doi.org/10.1088/1367-2630/17/7/075015}
	{\bibfield  {journal} {\bibinfo  {journal} {New J. Phys.}\ }\textbf {\bibinfo
			{volume} {17}},\ \bibinfo {pages} {075015} (\bibinfo {year}
		{2015})}\BibitemShut {NoStop}%
	\bibitem [{\citenamefont {Campaioli}\ \emph {et~al.}(2017)\citenamefont
		{Campaioli}, \citenamefont {Pollock}, \citenamefont {Binder}, \citenamefont
		{C\'eleri}, \citenamefont {Goold}, \citenamefont {Vinjanampathy},\ and\
		\citenamefont {Modi}}]{PhysRevLett.118.150601}%
	\BibitemOpen
	\bibfield  {author} {\bibinfo {author} {\bibfnamefont {F.}~\bibnamefont
			{Campaioli}}, \bibinfo {author} {\bibfnamefont {F.~A.}\ \bibnamefont
			{Pollock}}, \bibinfo {author} {\bibfnamefont {F.~C.}\ \bibnamefont {Binder}},
		\bibinfo {author} {\bibfnamefont {L.}~\bibnamefont {C\'eleri}}, \bibinfo
		{author} {\bibfnamefont {J.}~\bibnamefont {Goold}}, \bibinfo {author}
		{\bibfnamefont {S.}~\bibnamefont {Vinjanampathy}},\ and\ \bibinfo {author}
		{\bibfnamefont {K.}~\bibnamefont {Modi}},\ }\bibfield  {title} {\bibinfo
		{title} {Enhancing the charging power of quantum batteries},\ }\href
	{https://doi.org/10.1103/PhysRevLett.118.150601} {\bibfield  {journal}
		{\bibinfo  {journal} {Phys. Rev. Lett.}\ }\textbf {\bibinfo {volume} {118}},\
		\bibinfo {pages} {150601} (\bibinfo {year} {2017})}\BibitemShut {NoStop}%
	\bibitem [{\citenamefont {Juli\`a-Farr\'e}\ \emph {et~al.}(2020)\citenamefont
		{Juli\`a-Farr\'e}, \citenamefont {Salamon}, \citenamefont {Riera},
		\citenamefont {Bera},\ and\ \citenamefont
		{Lewenstein}}]{Julia-Farre:2020jke}%
	\BibitemOpen
	\bibfield  {author} {\bibinfo {author} {\bibfnamefont {S.}~\bibnamefont
			{Juli\`a-Farr\'e}}, \bibinfo {author} {\bibfnamefont {T.}~\bibnamefont
			{Salamon}}, \bibinfo {author} {\bibfnamefont {A.}~\bibnamefont {Riera}},
		\bibinfo {author} {\bibfnamefont {M.~N.}\ \bibnamefont {Bera}},\ and\
		\bibinfo {author} {\bibfnamefont {M.}~\bibnamefont {Lewenstein}},\ }\bibfield
	{title} {\bibinfo {title} {{Bounds on the capacity and power of quantum
				batteries}},\ }\href {https://doi.org/10.1103/PhysRevResearch.2.023113}
	{\bibfield  {journal} {\bibinfo  {journal} {Phys. Rev. Res.}\ }\textbf
		{\bibinfo {volume} {2}},\ \bibinfo {pages} {023113} (\bibinfo {year}
		{2020})}\BibitemShut {NoStop}%
	\bibitem [{\citenamefont {Ferraro}\ \emph {et~al.}(2018)\citenamefont
		{Ferraro}, \citenamefont {Campisi}, \citenamefont {Andolina}, \citenamefont
		{Pellegrini},\ and\ \citenamefont {Polini}}]{Ferraro:2018tum}%
	\BibitemOpen
	\bibfield  {author} {\bibinfo {author} {\bibfnamefont {D.}~\bibnamefont
			{Ferraro}}, \bibinfo {author} {\bibfnamefont {M.}~\bibnamefont {Campisi}},
		\bibinfo {author} {\bibfnamefont {G.~M.}\ \bibnamefont {Andolina}}, \bibinfo
		{author} {\bibfnamefont {V.}~\bibnamefont {Pellegrini}},\ and\ \bibinfo
		{author} {\bibfnamefont {M.}~\bibnamefont {Polini}},\ }\bibfield  {title}
	{\bibinfo {title} {{High-Power Collective Charging of a Solid-State Quantum
				Battery}},\ }\href {https://doi.org/10.1103/PhysRevLett.120.117702}
	{\bibfield  {journal} {\bibinfo  {journal} {Phys. Rev. Lett.}\ }\textbf
		{\bibinfo {volume} {120}},\ \bibinfo {pages} {117702} (\bibinfo {year}
		{2018})}\BibitemShut {NoStop}%
	\bibitem [{\citenamefont {Le}\ \emph {et~al.}(2018)\citenamefont {Le},
		\citenamefont {Levinsen}, \citenamefont {Modi}, \citenamefont {Parish},\ and\
		\citenamefont {Pollock}}]{Le:2018fet}%
	\BibitemOpen
	\bibfield  {author} {\bibinfo {author} {\bibfnamefont {T.~P.}\ \bibnamefont
			{Le}}, \bibinfo {author} {\bibfnamefont {J.}~\bibnamefont {Levinsen}},
		\bibinfo {author} {\bibfnamefont {K.}~\bibnamefont {Modi}}, \bibinfo {author}
		{\bibfnamefont {M.~M.}\ \bibnamefont {Parish}},\ and\ \bibinfo {author}
		{\bibfnamefont {F.~A.}\ \bibnamefont {Pollock}},\ }\bibfield  {title}
	{\bibinfo {title} {{Spin-chain model of a many-body quantum battery}},\
	}\href {https://doi.org/10.1103/PhysRevA.97.022106} {\bibfield  {journal}
		{\bibinfo  {journal} {Phys. Rev. A}\ }\textbf {\bibinfo {volume} {97}},\
		\bibinfo {pages} {022106} (\bibinfo {year} {2018})}\BibitemShut {NoStop}%
	\bibitem [{\citenamefont {Grazi}\ \emph
		{et~al.}(2025{\natexlab{a}})\citenamefont {Grazi}, \citenamefont {Cavaliere},
		\citenamefont {Sassetti}, \citenamefont {Ferraro},\ and\ \citenamefont
		{Ziani}}]{Grazi:2025tlx}%
	\BibitemOpen
	\bibfield  {author} {\bibinfo {author} {\bibfnamefont {R.}~\bibnamefont
			{Grazi}}, \bibinfo {author} {\bibfnamefont {F.}~\bibnamefont {Cavaliere}},
		\bibinfo {author} {\bibfnamefont {M.}~\bibnamefont {Sassetti}}, \bibinfo
		{author} {\bibfnamefont {D.}~\bibnamefont {Ferraro}},\ and\ \bibinfo {author}
		{\bibfnamefont {N.~T.}\ \bibnamefont {Ziani}},\ }\bibfield  {title} {\bibinfo
		{title} {{Charging free fermion quantum batteries}},\ }\href
	{https://doi.org/10.1016/j.chaos.2025.116383} {\bibfield  {journal} {\bibinfo
			{journal} {Chaos Solitons Fractals}\ }\textbf {\bibinfo {volume} {196}},\
		\bibinfo {pages} {116383} (\bibinfo {year} {2025}{\natexlab{a}})},\ \Eprint
	{https://arxiv.org/abs/2503.20424} {arXiv:2503.20424 [quant-ph]} \BibitemShut
	{NoStop}%
	\bibitem [{\citenamefont {K}\ and\ \citenamefont {Murugesh}(2025)}]{K:2024sik}%
	\BibitemOpen
	\bibfield  {author} {\bibinfo {author} {\bibfnamefont {S.~R.}\ \bibnamefont
			{K}}\ and\ \bibinfo {author} {\bibfnamefont {S.}~\bibnamefont {Murugesh}},\
	}\bibfield  {title} {\bibinfo {title} {{Effect of DM Interaction in the
				charging process of a Heisenberg spin chain quantum battery}},\ }\href
	{https://doi.org/10.1088/1402-4896/ad95c5} {\bibfield  {journal} {\bibinfo
			{journal} {Phys. Scripta}\ }\textbf {\bibinfo {volume} {100}},\ \bibinfo
		{pages} {015106} (\bibinfo {year} {2025})}\BibitemShut {NoStop}%
	\bibitem [{\citenamefont {Grazi}\ \emph {et~al.}(2024)\citenamefont {Grazi},
		\citenamefont {Shaikh}, \citenamefont {Sassetti}, \citenamefont {Ziani},\
		and\ \citenamefont {Ferraro}}]{Grazi:2024kyr}%
	\BibitemOpen
	\bibfield  {author} {\bibinfo {author} {\bibfnamefont {R.}~\bibnamefont
			{Grazi}}, \bibinfo {author} {\bibfnamefont {D.~S.}\ \bibnamefont {Shaikh}},
		\bibinfo {author} {\bibfnamefont {M.}~\bibnamefont {Sassetti}}, \bibinfo
		{author} {\bibfnamefont {N.~T.}\ \bibnamefont {Ziani}},\ and\ \bibinfo
		{author} {\bibfnamefont {D.}~\bibnamefont {Ferraro}},\ }\bibfield  {title}
	{\bibinfo {title} {{Controlling Energy Storage Crossing Quantum Phase
				Transitions in an Integrable Spin Quantum Battery}},\ }\href
	{https://doi.org/10.1103/PhysRevLett.133.197001} {\bibfield  {journal}
		{\bibinfo  {journal} {Phys. Rev. Lett.}\ }\textbf {\bibinfo {volume} {133}},\
		\bibinfo {pages} {197001} (\bibinfo {year} {2024})},\ \Eprint
	{https://arxiv.org/abs/2402.09169} {arXiv:2402.09169 [quant-ph]} \BibitemShut
	{NoStop}%
	\bibitem [{\citenamefont {Grazi}\ \emph
		{et~al.}(2025{\natexlab{b}})\citenamefont {Grazi}, \citenamefont {Cavaliere},
		\citenamefont {Ziani},\ and\ \citenamefont {Ferraro}}]{Grazi:2025gxl}%
	\BibitemOpen
	\bibfield  {author} {\bibinfo {author} {\bibfnamefont {R.}~\bibnamefont
			{Grazi}}, \bibinfo {author} {\bibfnamefont {F.}~\bibnamefont {Cavaliere}},
		\bibinfo {author} {\bibfnamefont {N.~T.}\ \bibnamefont {Ziani}},\ and\
		\bibinfo {author} {\bibfnamefont {D.}~\bibnamefont {Ferraro}},\ }\bibfield
	{title} {\bibinfo {title} {{Charging a Dimerized Quantum XY Chain}},\ }\href
	{https://doi.org/10.3390/sym17020220} {\bibfield  {journal} {\bibinfo
			{journal} {Symmetry}\ }\textbf {\bibinfo {volume} {17}},\ \bibinfo {pages}
		{220} (\bibinfo {year} {2025}{\natexlab{b}})},\ \Eprint
	{https://arxiv.org/abs/2502.06503} {arXiv:2502.06503 [quant-ph]} \BibitemShut
	{NoStop}%
	\bibitem [{\citenamefont {Divi}\ \emph {et~al.}(2025)\citenamefont {Divi},
		\citenamefont {Murugan},\ and\ \citenamefont {Rosa}}]{Divi:2024mup}%
	\BibitemOpen
	\bibfield  {author} {\bibinfo {author} {\bibfnamefont {F.}~\bibnamefont
			{Divi}}, \bibinfo {author} {\bibfnamefont {J.}~\bibnamefont {Murugan}},\ and\
		\bibinfo {author} {\bibfnamefont {D.}~\bibnamefont {Rosa}},\ }\bibfield
	{title} {\bibinfo {title} {{Sachdev-Ye-Kitaev charging advantage as a random
				walk on graphs}},\ }\href {https://doi.org/10.1103/PhysRevB.111.075138}
	{\bibfield  {journal} {\bibinfo  {journal} {Phys. Rev. B}\ }\textbf {\bibinfo
			{volume} {111}},\ \bibinfo {pages} {075138} (\bibinfo {year} {2025})},\
	\Eprint {https://arxiv.org/abs/2412.04560} {arXiv:2412.04560 [quant-ph]}
	\BibitemShut {NoStop}%
	\bibitem [{\citenamefont {Rossini}\ \emph {et~al.}(2020)\citenamefont
		{Rossini}, \citenamefont {Andolina}, \citenamefont {Rosa}, \citenamefont
		{Carrega},\ and\ \citenamefont {Polini}}]{Rossini:2019nfu}%
	\BibitemOpen
	\bibfield  {author} {\bibinfo {author} {\bibfnamefont {D.}~\bibnamefont
			{Rossini}}, \bibinfo {author} {\bibfnamefont {G.~M.}\ \bibnamefont
			{Andolina}}, \bibinfo {author} {\bibfnamefont {D.}~\bibnamefont {Rosa}},
		\bibinfo {author} {\bibfnamefont {M.}~\bibnamefont {Carrega}},\ and\ \bibinfo
		{author} {\bibfnamefont {M.}~\bibnamefont {Polini}},\ }\bibfield  {title}
	{\bibinfo {title} {{Quantum advantage in the charging process of
				Sachdev-Ye-Kitaev batteries}},\ }\href
	{https://doi.org/10.1103/PhysRevLett.125.236402} {\bibfield  {journal}
		{\bibinfo  {journal} {Phys. Rev. Lett.}\ }\textbf {\bibinfo {volume} {125}},\
		\bibinfo {pages} {236402} (\bibinfo {year} {2020})},\ \Eprint
	{https://arxiv.org/abs/1912.07234} {arXiv:1912.07234 [cond-mat.str-el]}
	\BibitemShut {NoStop}%
	\bibitem [{\citenamefont {Rosa}\ \emph {et~al.}(2020)\citenamefont {Rosa},
		\citenamefont {Rossini}, \citenamefont {Andolina}, \citenamefont {Polini},\
		and\ \citenamefont {Carrega}}]{Rosa:2019jin}%
	\BibitemOpen
	\bibfield  {author} {\bibinfo {author} {\bibfnamefont {D.}~\bibnamefont
			{Rosa}}, \bibinfo {author} {\bibfnamefont {D.}~\bibnamefont {Rossini}},
		\bibinfo {author} {\bibfnamefont {G.~M.}\ \bibnamefont {Andolina}}, \bibinfo
		{author} {\bibfnamefont {M.}~\bibnamefont {Polini}},\ and\ \bibinfo {author}
		{\bibfnamefont {M.}~\bibnamefont {Carrega}},\ }\bibfield  {title} {\bibinfo
		{title} {{Ultra-stable charging of fast-scrambling SYK quantum batteries}},\
	}\href {https://doi.org/10.1007/JHEP11(2020)067} {\bibfield  {journal}
		{\bibinfo  {journal} {JHEP}\ }\textbf {\bibinfo {volume} {11}},\ \bibinfo
		{pages} {067}},\ \Eprint {https://arxiv.org/abs/1912.07247} {arXiv:1912.07247
		[cond-mat.str-el]} \BibitemShut {NoStop}%
	\bibitem [{\citenamefont {Romero}\ \emph {et~al.}(2025)\citenamefont {Romero},
		\citenamefont {Ding}, \citenamefont {Chen},\ and\ \citenamefont
		{Ban}}]{Romero:2024wgt}%
	\BibitemOpen
	\bibfield  {author} {\bibinfo {author} {\bibfnamefont {S.~V.}\ \bibnamefont
			{Romero}}, \bibinfo {author} {\bibfnamefont {Y.}~\bibnamefont {Ding}},
		\bibinfo {author} {\bibfnamefont {X.}~\bibnamefont {Chen}},\ and\ \bibinfo
		{author} {\bibfnamefont {Y.}~\bibnamefont {Ban}},\ }\bibfield  {title}
	{\bibinfo {title} {{Scrambling in the charging of quantum batteries}},\
	}\href {https://doi.org/10.1007/JHEP05(2025)021} {\bibfield  {journal}
		{\bibinfo  {journal} {JHEP}\ }\textbf {\bibinfo {volume} {05}},\ \bibinfo
		{pages} {021}},\ \Eprint {https://arxiv.org/abs/2409.10590} {arXiv:2409.10590
		[quant-ph]} \BibitemShut {NoStop}%
	\bibitem [{\citenamefont {Wang}\ \emph {et~al.}(2024)\citenamefont {Wang},
		\citenamefont {Liu}, \citenamefont {Wu}, \citenamefont {Fan},\ and\
		\citenamefont {Liu}}]{Wang:2024ogv}%
	\BibitemOpen
	\bibfield  {author} {\bibinfo {author} {\bibfnamefont {L.}~\bibnamefont
			{Wang}}, \bibinfo {author} {\bibfnamefont {S.-Q.}\ \bibnamefont {Liu}},
		\bibinfo {author} {\bibfnamefont {F.-l.}\ \bibnamefont {Wu}}, \bibinfo
		{author} {\bibfnamefont {H.}~\bibnamefont {Fan}},\ and\ \bibinfo {author}
		{\bibfnamefont {S.-Y.}\ \bibnamefont {Liu}},\ }\bibfield  {title} {\bibinfo
		{title} {{Cavity-optomechanical quantum battery}},\ }\href
	{https://doi.org/10.1103/PhysRevA.110.062204} {\bibfield  {journal} {\bibinfo
			{journal} {Phys. Rev. A}\ }\textbf {\bibinfo {volume} {110}},\ \bibinfo
		{pages} {062204} (\bibinfo {year} {2024})}\BibitemShut {NoStop}%
	\bibitem [{\citenamefont {Hu}\ \emph {et~al.}(2025{\natexlab{b}})\citenamefont
		{Hu}, \citenamefont {Gao},\ and\ \citenamefont {Fan}}]{Hu:2025aod}%
	\BibitemOpen
	\bibfield  {author} {\bibinfo {author} {\bibfnamefont {M.-L.}\ \bibnamefont
			{Hu}}, \bibinfo {author} {\bibfnamefont {T.}~\bibnamefont {Gao}},\ and\
		\bibinfo {author} {\bibfnamefont {H.}~\bibnamefont {Fan}},\ }\bibfield
	{title} {\bibinfo {title} {{Efficient wireless charging of a quantum
				battery}},\ }\href {https://doi.org/10.1103/PhysRevA.111.042216} {\bibfield
		{journal} {\bibinfo  {journal} {Phys. Rev. A}\ }\textbf {\bibinfo {volume}
			{111}},\ \bibinfo {pages} {042216} (\bibinfo {year} {2025}{\natexlab{b}})},\
	\Eprint {https://arxiv.org/abs/2501.08843} {arXiv:2501.08843 [quant-ph]}
	\BibitemShut {NoStop}%
	\bibitem [{\citenamefont {Dou}\ \emph {et~al.}(2022)\citenamefont {Dou},
		\citenamefont {Zhou},\ and\ \citenamefont {Sun}}]{PhysRevA.106.032212}%
	\BibitemOpen
	\bibfield  {author} {\bibinfo {author} {\bibfnamefont {F.-Q.}\ \bibnamefont
			{Dou}}, \bibinfo {author} {\bibfnamefont {H.}~\bibnamefont {Zhou}},\ and\
		\bibinfo {author} {\bibfnamefont {J.-A.}\ \bibnamefont {Sun}},\ }\bibfield
	{title} {\bibinfo {title} {Cavity heisenberg-spin-chain quantum battery},\
	}\href {https://doi.org/10.1103/PhysRevA.106.032212} {\bibfield  {journal}
		{\bibinfo  {journal} {Phys. Rev. A}\ }\textbf {\bibinfo {volume} {106}},\
		\bibinfo {pages} {032212} (\bibinfo {year} {2022})}\BibitemShut {NoStop}%
	\bibitem [{\citenamefont {Zhao}\ \emph
		{et~al.}(2025{\natexlab{a}})\citenamefont {Zhao}, \citenamefont {Zhao},\ and\
		\citenamefont {Zhuang}}]{Zhao:2025iki}%
	\BibitemOpen
	\bibfield  {author} {\bibinfo {author} {\bibfnamefont {S.-C.}\ \bibnamefont
			{Zhao}}, \bibinfo {author} {\bibfnamefont {Z.-R.}\ \bibnamefont {Zhao}},\
		and\ \bibinfo {author} {\bibfnamefont {N.-Y.}\ \bibnamefont {Zhuang}},\
	}\bibfield  {title} {\bibinfo {title} {{Non-Markovian N-spin chain quantum
				battery in thermal charging process}},\ }\href
	{https://doi.org/10.1103/xqtv-qbyk} {\bibfield  {journal} {\bibinfo
			{journal} {Phys. Rev. E}\ }\textbf {\bibinfo {volume} {112}},\ \bibinfo
		{pages} {024129} (\bibinfo {year} {2025}{\natexlab{a}})},\ \Eprint
	{https://arxiv.org/abs/2504.05348} {arXiv:2504.05348 [quant-ph]} \BibitemShut
	{NoStop}%
	\bibitem [{\citenamefont {Sun}\ \emph {et~al.}(2024)\citenamefont {Sun},
		\citenamefont {Zhou},\ and\ \citenamefont {Dou}}]{Sun2024CavityHeisenbergSC}%
	\BibitemOpen
	\bibfield  {author} {\bibinfo {author} {\bibfnamefont {P.-Y.}\ \bibnamefont
			{Sun}}, \bibinfo {author} {\bibfnamefont {H.}~\bibnamefont {Zhou}},\ and\
		\bibinfo {author} {\bibfnamefont {F.-Q.}\ \bibnamefont {Dou}},\ }\bibfield
	{title} {\bibinfo {title} {Cavity-heisenberg spin-\$j\$ chain quantum battery
			and reinforcement learning optimization}\ }(\bibinfo {year}
	{2024})\BibitemShut {NoStop}%
	\bibitem [{\citenamefont {Zhang}\ \emph {et~al.}(2024)\citenamefont {Zhang},
		\citenamefont {Ma}, \citenamefont {Jiang}, \citenamefont {Yu}, \citenamefont
		{Jin},\ and\ \citenamefont {Chen}}]{Zhang:2024bva}%
	\BibitemOpen
	\bibfield  {author} {\bibinfo {author} {\bibfnamefont {D.}~\bibnamefont
			{Zhang}}, \bibinfo {author} {\bibfnamefont {S.}~\bibnamefont {Ma}}, \bibinfo
		{author} {\bibfnamefont {Y.}~\bibnamefont {Jiang}}, \bibinfo {author}
		{\bibfnamefont {Y.}~\bibnamefont {Yu}}, \bibinfo {author} {\bibfnamefont
			{G.}~\bibnamefont {Jin}},\ and\ \bibinfo {author} {\bibfnamefont
			{A.}~\bibnamefont {Chen}},\ }\bibfield  {title} {\bibinfo {title} {{Quantum
				battery with interactive atomic collective charging}},\ }\href
	{https://doi.org/10.1103/PhysRevA.110.032211} {\bibfield  {journal} {\bibinfo
			{journal} {Phys. Rev. A}\ }\textbf {\bibinfo {volume} {110}},\ \bibinfo
		{pages} {032211} (\bibinfo {year} {2024})}\BibitemShut {NoStop}%
	\bibitem [{\citenamefont {Zhang}\ \emph {et~al.}(2025)\citenamefont {Zhang},
		\citenamefont {Ma}, \citenamefont {Jiang}, \citenamefont {Yu}, \citenamefont
		{Jin},\ and\ \citenamefont {Chen}}]{Zhang:2024hvm}%
	\BibitemOpen
	\bibfield  {author} {\bibinfo {author} {\bibfnamefont {D.}~\bibnamefont
			{Zhang}}, \bibinfo {author} {\bibfnamefont {S.}~\bibnamefont {Ma}}, \bibinfo
		{author} {\bibfnamefont {Y.}~\bibnamefont {Jiang}}, \bibinfo {author}
		{\bibfnamefont {Y.}~\bibnamefont {Yu}}, \bibinfo {author} {\bibfnamefont
			{G.}~\bibnamefont {Jin}},\ and\ \bibinfo {author} {\bibfnamefont
			{A.}~\bibnamefont {Chen}},\ }\bibfield  {title} {\bibinfo {title}
		{{Entanglement and Steering in Quantum Batteries}},\ }\href
	{https://doi.org/10.1002/qute.202500243} {\bibfield  {journal} {\bibinfo
			{journal} {Adv. Quantum Technol.}\ }\textbf {\bibinfo {volume} {8}},\
		\bibinfo {pages} {e2500243} (\bibinfo {year} {2025})},\ \Eprint
	{https://arxiv.org/abs/2406.06373} {arXiv:2406.06373 [quant-ph]} \BibitemShut
	{NoStop}%
	\bibitem [{\citenamefont {Konar}\ \emph {et~al.}(2024)\citenamefont {Konar},
		\citenamefont {Patra}, \citenamefont {Gupta}, \citenamefont {Ghosh},\ and\
		\citenamefont {De}}]{Konar:2022myn}%
	\BibitemOpen
	\bibfield  {author} {\bibinfo {author} {\bibfnamefont {T.~K.}\ \bibnamefont
			{Konar}}, \bibinfo {author} {\bibfnamefont {A.}~\bibnamefont {Patra}},
		\bibinfo {author} {\bibfnamefont {R.}~\bibnamefont {Gupta}}, \bibinfo
		{author} {\bibfnamefont {S.}~\bibnamefont {Ghosh}},\ and\ \bibinfo {author}
		{\bibfnamefont {A.~S.}\ \bibnamefont {De}},\ }\bibfield  {title} {\bibinfo
		{title} {{Multimode advantage in continuous-variable quantum batteries}},\
	}\href {https://doi.org/10.1103/PhysRevA.110.022226} {\bibfield  {journal}
		{\bibinfo  {journal} {Phys. Rev. A}\ }\textbf {\bibinfo {volume} {110}},\
		\bibinfo {pages} {022226} (\bibinfo {year} {2024})},\ \Eprint
	{https://arxiv.org/abs/2210.16528} {arXiv:2210.16528 [quant-ph]} \BibitemShut
	{NoStop}%
	\bibitem [{\citenamefont {Downing}\ and\ \citenamefont
		{Ukhtary}(2024)}]{Downing:2024qzh}%
	\BibitemOpen
	\bibfield  {author} {\bibinfo {author} {\bibfnamefont {C.~A.}\ \bibnamefont
			{Downing}}\ and\ \bibinfo {author} {\bibfnamefont {M.~S.}\ \bibnamefont
			{Ukhtary}},\ }\bibfield  {title} {\bibinfo {title} {{Energetics of a pulsed
				quantum battery}},\ }\href {https://doi.org/10.1209/0295-5075/ad2e79}
	{\bibfield  {journal} {\bibinfo  {journal} {EPL}\ }\textbf {\bibinfo {volume}
			{146}},\ \bibinfo {pages} {10001} (\bibinfo {year} {2024})},\ \Eprint
	{https://arxiv.org/abs/2403.20155} {arXiv:2403.20155 [quant-ph]} \BibitemShut
	{NoStop}%
	\bibitem [{\citenamefont {Downing}\ and\ \citenamefont
		{Ukhtary}(2025)}]{Downing:2025hfg}%
	\BibitemOpen
	\bibfield  {author} {\bibinfo {author} {\bibfnamefont {C.~A.}\ \bibnamefont
			{Downing}}\ and\ \bibinfo {author} {\bibfnamefont {M.~S.}\ \bibnamefont
			{Ukhtary}},\ }\bibfield  {title} {\bibinfo {title} {{Energy storage in a
				continuous-variable quantum battery with nonlinear coupling}},\ }\href
	{https://doi.org/10.1103/73zl-yn4h} {\bibfield  {journal} {\bibinfo
			{journal} {Phys. Rev. E}\ }\textbf {\bibinfo {volume} {112}},\ \bibinfo
		{pages} {044143} (\bibinfo {year} {2025})},\ \Eprint
	{https://arxiv.org/abs/2510.21672} {arXiv:2510.21672 [quant-ph]} \BibitemShut
	{NoStop}%
	\bibitem [{\citenamefont {Tian}\ \emph {et~al.}(2025)\citenamefont {Tian},
		\citenamefont {Liu}, \citenamefont {Wang},\ and\ \citenamefont
		{Jing}}]{Tian:2024wby}%
	\BibitemOpen
	\bibfield  {author} {\bibinfo {author} {\bibfnamefont {Z.}~\bibnamefont
			{Tian}}, \bibinfo {author} {\bibfnamefont {X.}~\bibnamefont {Liu}}, \bibinfo
		{author} {\bibfnamefont {J.}~\bibnamefont {Wang}},\ and\ \bibinfo {author}
		{\bibfnamefont {J.}~\bibnamefont {Jing}},\ }\bibfield  {title} {\bibinfo
		{title} {{Dissipative dynamics of an open quantum battery in the BTZ
				spacetime}},\ }\href {https://doi.org/10.1007/JHEP04(2025)188} {\bibfield
		{journal} {\bibinfo  {journal} {JHEP}\ }\textbf {\bibinfo {volume} {04}},\
		\bibinfo {pages} {188}},\ \Eprint {https://arxiv.org/abs/2409.09259}
	{arXiv:2409.09259 [hep-th]} \BibitemShut {NoStop}%
	\bibitem [{\citenamefont {{Liu}}\ \emph {et~al.}(2025)\citenamefont {{Liu}},
		\citenamefont {{Tian}},\ and\ \citenamefont {{Wang}}}]{2025arXiv250607568L}%
	\BibitemOpen
	\bibfield  {author} {\bibinfo {author} {\bibfnamefont {X.}~\bibnamefont
			{{Liu}}}, \bibinfo {author} {\bibfnamefont {Z.}~\bibnamefont {{Tian}}},\ and\
		\bibinfo {author} {\bibfnamefont {J.}~\bibnamefont {{Wang}}},\ }\bibfield
	{title} {\bibinfo {title} {{Open quantum battery in the background of a
				three-dimensional rotating black hole}},\ }\href
	{https://doi.org/10.48550/arXiv.2506.07568} {\bibfield  {journal} {\bibinfo
			{journal} {arXiv e-prints}\ ,\ \bibinfo {eid} {arXiv:2506.07568}} (\bibinfo
		{year} {2025})},\ \Eprint {https://arxiv.org/abs/2506.07568}
	{arXiv:2506.07568 [gr-qc]} \BibitemShut {NoStop}%
	\bibitem [{\citenamefont {Hao}\ \emph {et~al.}(2023)\citenamefont {Hao},
		\citenamefont {Yan}, \citenamefont {Tan},\ and\ \citenamefont
		{Wu}}]{Hao:2023ndo}%
	\BibitemOpen
	\bibfield  {author} {\bibinfo {author} {\bibfnamefont {X.}~\bibnamefont
			{Hao}}, \bibinfo {author} {\bibfnamefont {K.}~\bibnamefont {Yan}}, \bibinfo
		{author} {\bibfnamefont {J.}~\bibnamefont {Tan}},\ and\ \bibinfo {author}
		{\bibfnamefont {Q.-Y.}\ \bibnamefont {Wu}},\ }\bibfield  {title} {\bibinfo
		{title} {{Quantum work extraction of an accelerated Unruh-DeWitt battery in
				relativistic motion}},\ }\href {https://doi.org/10.1103/PhysRevA.107.012207}
	{\bibfield  {journal} {\bibinfo  {journal} {Phys. Rev. A}\ }\textbf {\bibinfo
			{volume} {107}},\ \bibinfo {pages} {012207} (\bibinfo {year}
		{2023})}\BibitemShut {NoStop}%
	\bibitem [{\citenamefont {Liu}\ \emph {et~al.}(2025{\natexlab{a}})\citenamefont
		{Liu}, \citenamefont {Tian},\ and\ \citenamefont {Jing}}]{Liu:2025bzv}%
	\BibitemOpen
	\bibfield  {author} {\bibinfo {author} {\bibfnamefont {X.}~\bibnamefont
			{Liu}}, \bibinfo {author} {\bibfnamefont {Z.}~\bibnamefont {Tian}},\ and\
		\bibinfo {author} {\bibfnamefont {J.}~\bibnamefont {Jing}},\ }\bibfield
	{title} {\bibinfo {title} {{Dissipation suppression for an Unruh-DeWitt
				battery with a reflecting boundary}},\ }\href
	{https://doi.org/10.1007/s11433-025-2723-7} {\bibfield  {journal} {\bibinfo
			{journal} {Sci. China Phys. Mech. Astron.}\ }\textbf {\bibinfo {volume}
			{68}},\ \bibinfo {pages} {100412} (\bibinfo {year} {2025}{\natexlab{a}})},\
	\Eprint {https://arxiv.org/abs/2509.00875} {arXiv:2509.00875 [hep-th]}
	\BibitemShut {NoStop}%
	\bibitem [{\citenamefont {Chen}\ \emph {et~al.}(2025)\citenamefont {Chen},
		\citenamefont {Zhang}, \citenamefont {Ren},\ and\ \citenamefont
		{Hao}}]{Chen:2025xaf}%
	\BibitemOpen
	\bibfield  {author} {\bibinfo {author} {\bibfnamefont {Y.}~\bibnamefont
			{Chen}}, \bibinfo {author} {\bibfnamefont {W.-W.}\ \bibnamefont {Zhang}},
		\bibinfo {author} {\bibfnamefont {T.-X.}\ \bibnamefont {Ren}},\ and\ \bibinfo
		{author} {\bibfnamefont {X.}~\bibnamefont {Hao}},\ }\bibfield  {title}
	{\bibinfo {title} {{Quantum work extraction of a moving battery as a witness
				to Unruh thermality in high-dimensional spacetimes}},\ }\href
	{https://doi.org/10.1103/PhysRevD.111.065028} {\bibfield  {journal} {\bibinfo
			{journal} {Phys. Rev. D}\ }\textbf {\bibinfo {volume} {111}},\ \bibinfo
		{pages} {065028} (\bibinfo {year} {2025})},\ \Eprint
	{https://arxiv.org/abs/2503.02472} {arXiv:2503.02472 [hep-th]} \BibitemShut
	{NoStop}%
	\bibitem [{\citenamefont {Xie}\ \emph {et~al.}(2024)\citenamefont {Xie},
		\citenamefont {Zhu}, \citenamefont {Tan},\ and\ \citenamefont
		{Hao}}]{Xie:2024hwg}%
	\BibitemOpen
	\bibfield  {author} {\bibinfo {author} {\bibfnamefont {J.-L.}\ \bibnamefont
			{Xie}}, \bibinfo {author} {\bibfnamefont {C.-J.}\ \bibnamefont {Zhu}},
		\bibinfo {author} {\bibfnamefont {J.}~\bibnamefont {Tan}},\ and\ \bibinfo
		{author} {\bibfnamefont {X.}~\bibnamefont {Hao}},\ }\bibfield  {title}
	{\bibinfo {title} {{Weak measurements enhancing the quantum information
				facets of a driven Unruh{\textendash}DeWitt detector}},\ }\href
	{https://doi.org/10.3389/fphy.2024.1513241} {\bibfield  {journal} {\bibinfo
			{journal} {Front. in Phys.}\ }\textbf {\bibinfo {volume} {12}},\ \bibinfo
		{pages} {1513241} (\bibinfo {year} {2024})}\BibitemShut {NoStop}%
	\bibitem [{\citenamefont {Mukherjee}\ \emph {et~al.}(2024)\citenamefont
		{Mukherjee}, \citenamefont {Gangopadhyay},\ and\ \citenamefont
		{Majumdar}}]{Mukherjee2024EnhancementOA}%
	\BibitemOpen
	\bibfield  {author} {\bibinfo {author} {\bibfnamefont {A.}~\bibnamefont
			{Mukherjee}}, \bibinfo {author} {\bibfnamefont {S.}~\bibnamefont
			{Gangopadhyay}},\ and\ \bibinfo {author} {\bibfnamefont {A.~S.}\ \bibnamefont
			{Majumdar}},\ }\bibfield  {title} {\bibinfo {title} {Enhancement of an
			unruh-dewitt battery performance through quadratic environmental coupling}\
	}(\bibinfo {year} {2024})\BibitemShut {NoStop}%
	\bibitem [{\citenamefont {Campaioli}\ \emph {et~al.}(2024)\citenamefont
		{Campaioli}, \citenamefont {Gherardini}, \citenamefont {Quach}, \citenamefont
		{Polini},\ and\ \citenamefont {Andolina}}]{Campaioli:2023ndh}%
	\BibitemOpen
	\bibfield  {author} {\bibinfo {author} {\bibfnamefont {F.}~\bibnamefont
			{Campaioli}}, \bibinfo {author} {\bibfnamefont {S.}~\bibnamefont
			{Gherardini}}, \bibinfo {author} {\bibfnamefont {J.~Q.}\ \bibnamefont
			{Quach}}, \bibinfo {author} {\bibfnamefont {M.}~\bibnamefont {Polini}},\ and\
		\bibinfo {author} {\bibfnamefont {G.~M.}\ \bibnamefont {Andolina}},\
	}\bibfield  {title} {\bibinfo {title} {{Colloquium: Quantum batteries}},\
	}\href {https://doi.org/10.1103/RevModPhys.96.031001} {\bibfield  {journal}
		{\bibinfo  {journal} {Rev. Mod. Phys.}\ }\textbf {\bibinfo {volume} {96}},\
		\bibinfo {pages} {031001} (\bibinfo {year} {2024})},\ \Eprint
	{https://arxiv.org/abs/2308.02277} {arXiv:2308.02277 [quant-ph]} \BibitemShut
	{NoStop}%
	\bibitem [{\citenamefont {Hu}\ \emph {et~al.}(2022)\citenamefont {Hu} \emph
		{et~al.}}]{Hu:2021klf}%
	\BibitemOpen
	\bibfield  {author} {\bibinfo {author} {\bibfnamefont {C.-K.}\ \bibnamefont
			{Hu}} \emph {et~al.},\ }\bibfield  {title} {\bibinfo {title} {{Optimal
				charging of a superconducting quantum battery}},\ }\href
	{https://doi.org/10.1088/2058-9565/ac8444} {\bibfield  {journal} {\bibinfo
			{journal} {Quantum Sci. Technol.}\ }\textbf {\bibinfo {volume} {7}},\
		\bibinfo {pages} {045018} (\bibinfo {year} {2022})},\ \Eprint
	{https://arxiv.org/abs/2108.04298} {arXiv:2108.04298 [quant-ph]} \BibitemShut
	{NoStop}%
	\bibitem [{\citenamefont {Joshi}\ and\ \citenamefont
		{Mahesh}(2022)}]{Joshi:2021aee}%
	\BibitemOpen
	\bibfield  {author} {\bibinfo {author} {\bibfnamefont {J.}~\bibnamefont
			{Joshi}}\ and\ \bibinfo {author} {\bibfnamefont {T.~S.}\ \bibnamefont
			{Mahesh}},\ }\bibfield  {title} {\bibinfo {title} {{Experimental
				investigation of a quantum battery using star-topology NMR spin systems}},\
	}\href {https://doi.org/10.1103/PhysRevA.106.042601} {\bibfield  {journal}
		{\bibinfo  {journal} {Phys. Rev. A}\ }\textbf {\bibinfo {volume} {106}},\
		\bibinfo {pages} {042601} (\bibinfo {year} {2022})},\ \Eprint
	{https://arxiv.org/abs/2112.15437} {arXiv:2112.15437 [quant-ph]} \BibitemShut
	{NoStop}%
	\bibitem [{\citenamefont {Kobrin}\ \emph {et~al.}(2021)\citenamefont {Kobrin},
		\citenamefont {Yang}, \citenamefont {Kahanamoku-Meyer}, \citenamefont
		{Olund}, \citenamefont {Moore}, \citenamefont {Stanford},\ and\ \citenamefont
		{Yao}}]{Kobrin:2020xms}%
	\BibitemOpen
	\bibfield  {author} {\bibinfo {author} {\bibfnamefont {B.}~\bibnamefont
			{Kobrin}}, \bibinfo {author} {\bibfnamefont {Z.}~\bibnamefont {Yang}},
		\bibinfo {author} {\bibfnamefont {G.~D.}\ \bibnamefont {Kahanamoku-Meyer}},
		\bibinfo {author} {\bibfnamefont {C.~T.}\ \bibnamefont {Olund}}, \bibinfo
		{author} {\bibfnamefont {J.~E.}\ \bibnamefont {Moore}}, \bibinfo {author}
		{\bibfnamefont {D.}~\bibnamefont {Stanford}},\ and\ \bibinfo {author}
		{\bibfnamefont {N.~Y.}\ \bibnamefont {Yao}},\ }\bibfield  {title} {\bibinfo
		{title} {{Many-Body Chaos in the Sachdev-Ye-Kitaev Model}},\ }\href
	{https://doi.org/10.1103/PhysRevLett.126.030602} {\bibfield  {journal}
		{\bibinfo  {journal} {Phys. Rev. Lett.}\ }\textbf {\bibinfo {volume} {126}},\
		\bibinfo {pages} {030602} (\bibinfo {year} {2021})},\ \Eprint
	{https://arxiv.org/abs/2002.05725} {arXiv:2002.05725 [hep-th]} \BibitemShut
	{NoStop}%
	\bibitem [{\citenamefont {Gyhm}\ and\ \citenamefont
		{Fischer}(2023)}]{Gyhm:2023lhb}%
	\BibitemOpen
	\bibfield  {author} {\bibinfo {author} {\bibfnamefont {J.-Y.}\ \bibnamefont
			{Gyhm}}\ and\ \bibinfo {author} {\bibfnamefont {U.~R.}\ \bibnamefont
			{Fischer}},\ }\bibfield  {title} {\bibinfo {title} {{Beneficial and
				detrimental entanglement for quantum battery charging}},\ }\href
	{https://doi.org/10.1116/5.0184903} {\bibfield  {journal} {\bibinfo
			{journal} {AVS Quantum Sci.}\ }\textbf {\bibinfo {volume} {6}},\ \bibinfo
		{pages} {012001} (\bibinfo {year} {2023})},\ \Eprint
	{https://arxiv.org/abs/2303.07841} {arXiv:2303.07841 [quant-ph]} \BibitemShut
	{NoStop}%
	\bibitem [{\citenamefont {Erdman}\ \emph {et~al.}(2024)\citenamefont {Erdman},
		\citenamefont {Andolina}, \citenamefont {Giovannetti},\ and\ \citenamefont
		{No{\'e}}}]{Erdman:2022oyp}%
	\BibitemOpen
	\bibfield  {author} {\bibinfo {author} {\bibfnamefont {P.~A.}\ \bibnamefont
			{Erdman}}, \bibinfo {author} {\bibfnamefont {G.~M.}\ \bibnamefont
			{Andolina}}, \bibinfo {author} {\bibfnamefont {V.}~\bibnamefont
			{Giovannetti}},\ and\ \bibinfo {author} {\bibfnamefont {F.}~\bibnamefont
			{No{\'e}}},\ }\bibfield  {title} {\bibinfo {title} {{Reinforcement Learning
				Optimization of the Charging of a Dicke Quantum Battery}},\ }\href
	{https://doi.org/10.1103/PhysRevLett.133.243602} {\bibfield  {journal}
		{\bibinfo  {journal} {Phys. Rev. Lett.}\ }\textbf {\bibinfo {volume} {133}},\
		\bibinfo {pages} {243602} (\bibinfo {year} {2024})},\ \Eprint
	{https://arxiv.org/abs/2212.12397} {arXiv:2212.12397 [quant-ph]} \BibitemShut
	{NoStop}%
	\bibitem [{\citenamefont {Maldacena}\ \emph {et~al.}(2016)\citenamefont
		{Maldacena}, \citenamefont {Shenker},\ and\ \citenamefont
		{Stanford}}]{Maldacena:2015waa}%
	\BibitemOpen
	\bibfield  {author} {\bibinfo {author} {\bibfnamefont {J.}~\bibnamefont
			{Maldacena}}, \bibinfo {author} {\bibfnamefont {S.~H.}\ \bibnamefont
			{Shenker}},\ and\ \bibinfo {author} {\bibfnamefont {D.}~\bibnamefont
			{Stanford}},\ }\bibfield  {title} {\bibinfo {title} {{A bound on chaos}},\
	}\href {https://doi.org/10.1007/JHEP08(2016)106} {\bibfield  {journal}
		{\bibinfo  {journal} {JHEP}\ }\textbf {\bibinfo {volume} {08}},\ \bibinfo
		{pages} {106}},\ \Eprint {https://arxiv.org/abs/1503.01409} {arXiv:1503.01409
		[hep-th]} \BibitemShut {NoStop}%
	\bibitem [{\citenamefont {Shenker}\ and\ \citenamefont
		{Stanford}(2014)}]{Shenker:2013pqa}%
	\BibitemOpen
	\bibfield  {author} {\bibinfo {author} {\bibfnamefont {S.~H.}\ \bibnamefont
			{Shenker}}\ and\ \bibinfo {author} {\bibfnamefont {D.}~\bibnamefont
			{Stanford}},\ }\bibfield  {title} {\bibinfo {title} {{Black holes and the
				butterfly effect}},\ }\href {https://doi.org/10.1007/JHEP03(2014)067}
	{\bibfield  {journal} {\bibinfo  {journal} {JHEP}\ }\textbf {\bibinfo
			{volume} {03}},\ \bibinfo {pages} {067}},\ \Eprint
	{https://arxiv.org/abs/1306.0622} {arXiv:1306.0622 [hep-th]} \BibitemShut
	{NoStop}%
	\bibitem [{\citenamefont {Yang}\ \emph {et~al.}(2020)\citenamefont {Yang},
		\citenamefont {Liu}, \citenamefont {Zhu}, \citenamefont {Luo},\ and\
		\citenamefont {Cai}}]{Yang:2019kbb}%
	\BibitemOpen
	\bibfield  {author} {\bibinfo {author} {\bibfnamefont {R.-Q.}\ \bibnamefont
			{Yang}}, \bibinfo {author} {\bibfnamefont {H.}~\bibnamefont {Liu}}, \bibinfo
		{author} {\bibfnamefont {S.}~\bibnamefont {Zhu}}, \bibinfo {author}
		{\bibfnamefont {L.}~\bibnamefont {Luo}},\ and\ \bibinfo {author}
		{\bibfnamefont {R.-G.}\ \bibnamefont {Cai}},\ }\bibfield  {title} {\bibinfo
		{title} {{Simulating quantum field theory in curved spacetime with quantum
				many-body systems}},\ }\href
	{https://doi.org/10.1103/PhysRevResearch.2.023107} {\bibfield  {journal}
		{\bibinfo  {journal} {Phys. Rev. Res.}\ }\textbf {\bibinfo {volume} {2}},\
		\bibinfo {pages} {023107} (\bibinfo {year} {2020})},\ \Eprint
	{https://arxiv.org/abs/1906.01927} {arXiv:1906.01927 [gr-qc]} \BibitemShut
	{NoStop}%
	\bibitem [{\citenamefont {Liu}\ \emph {et~al.}(2025{\natexlab{b}})\citenamefont
		{Liu}, \citenamefont {Yang}, \citenamefont {Fan},\ and\ \citenamefont
		{Wang}}]{Liu:2024wqj}%
	\BibitemOpen
	\bibfield  {author} {\bibinfo {author} {\bibfnamefont {Z.}~\bibnamefont
			{Liu}}, \bibinfo {author} {\bibfnamefont {R.-Q.}\ \bibnamefont {Yang}},
		\bibinfo {author} {\bibfnamefont {H.}~\bibnamefont {Fan}},\ and\ \bibinfo
		{author} {\bibfnamefont {J.}~\bibnamefont {Wang}},\ }\bibfield  {title}
	{\bibinfo {title} {{Simulation of the massless Dirac field in 1+1D curved
				spacetime}},\ }\href {https://doi.org/10.1007/s11433-025-2696-3} {\bibfield
		{journal} {\bibinfo  {journal} {Sci. China Phys. Mech. Astron.}\ }\textbf
		{\bibinfo {volume} {68}},\ \bibinfo {pages} {290411} (\bibinfo {year}
		{2025}{\natexlab{b}})},\ \Eprint {https://arxiv.org/abs/2411.15695}
	{arXiv:2411.15695 [gr-qc]} \BibitemShut {NoStop}%
	\bibitem [{\citenamefont {Shi}\ \emph {et~al.}(2023)\citenamefont {Shi} \emph
		{et~al.}}]{Shi:2021nkx}%
	\BibitemOpen
	\bibfield  {author} {\bibinfo {author} {\bibfnamefont {Y.-H.}\ \bibnamefont
			{Shi}} \emph {et~al.},\ }\bibfield  {title} {\bibinfo {title} {{Quantum
				simulation of Hawking radiation and curved spacetime with a superconducting
				on-chip black hole}},\ }\href {https://doi.org/10.1038/s41467-023-39064-6}
	{\bibfield  {journal} {\bibinfo  {journal} {Nature Commun.}\ }\textbf
		{\bibinfo {volume} {14}},\ \bibinfo {pages} {3263} (\bibinfo {year}
		{2023})},\ \Eprint {https://arxiv.org/abs/2111.11092} {arXiv:2111.11092
		[quant-ph]} \BibitemShut {NoStop}%
	\bibitem [{\citenamefont {Houck}\ \emph {et~al.}(2012)\citenamefont {Houck},
		\citenamefont {T{\"u}reci},\ and\ \citenamefont {Koch}}]{Houck:2012hbf}%
	\BibitemOpen
	\bibfield  {author} {\bibinfo {author} {\bibfnamefont {A.~A.}\ \bibnamefont
			{Houck}}, \bibinfo {author} {\bibfnamefont {H.~E.}\ \bibnamefont
			{T{\"u}reci}},\ and\ \bibinfo {author} {\bibfnamefont {J.}~\bibnamefont
			{Koch}},\ }\bibfield  {title} {\bibinfo {title} {{On-chip quantum simulation
				with superconducting circuits}},\ }\href {https://doi.org/10.1038/nphys2251}
	{\bibfield  {journal} {\bibinfo  {journal} {Nature Phys.}\ }\textbf {\bibinfo
			{volume} {8}},\ \bibinfo {pages} {292} (\bibinfo {year} {2012})}\BibitemShut
	{NoStop}%
	\bibitem [{\citenamefont {Kinoshita}\ \emph
		{et~al.}(2025{\natexlab{a}})\citenamefont {Kinoshita}, \citenamefont
		{Murata}, \citenamefont {Yamamoto},\ and\ \citenamefont
		{Yoshii}}]{Kinoshita:2024ahu}%
	\BibitemOpen
	\bibfield  {author} {\bibinfo {author} {\bibfnamefont {S.}~\bibnamefont
			{Kinoshita}}, \bibinfo {author} {\bibfnamefont {K.}~\bibnamefont {Murata}},
		\bibinfo {author} {\bibfnamefont {D.}~\bibnamefont {Yamamoto}},\ and\
		\bibinfo {author} {\bibfnamefont {R.}~\bibnamefont {Yoshii}},\ }\bibfield
	{title} {\bibinfo {title} {{Spin systems as quantum simulators of quantum
				field theories in curved spacetimes}},\ }\href
	{https://doi.org/10.1103/PhysRevResearch.7.023197} {\bibfield  {journal}
		{\bibinfo  {journal} {Phys. Rev. Res.}\ }\textbf {\bibinfo {volume} {7}},\
		\bibinfo {pages} {023197} (\bibinfo {year} {2025}{\natexlab{a}})},\ \Eprint
	{https://arxiv.org/abs/2410.07587} {arXiv:2410.07587 [hep-th]} \BibitemShut
	{NoStop}%
	\bibitem [{\citenamefont {Deger}\ \emph {et~al.}(2023)\citenamefont {Deger},
		\citenamefont {Horner},\ and\ \citenamefont {Pachos}}]{Deger:2022qob}%
	\BibitemOpen
	\bibfield  {author} {\bibinfo {author} {\bibfnamefont {A.}~\bibnamefont
			{Deger}}, \bibinfo {author} {\bibfnamefont {M.~D.}\ \bibnamefont {Horner}},\
		and\ \bibinfo {author} {\bibfnamefont {J.~K.}\ \bibnamefont {Pachos}},\
	}\bibfield  {title} {\bibinfo {title} {{AdS/CFT correspondence with a
				three-dimensional black hole simulator}},\ }\href
	{https://doi.org/10.1103/PhysRevB.108.155124} {\bibfield  {journal} {\bibinfo
			{journal} {Phys. Rev. B}\ }\textbf {\bibinfo {volume} {108}},\ \bibinfo
		{pages} {155124} (\bibinfo {year} {2023})},\ \Eprint
	{https://arxiv.org/abs/2211.15305} {arXiv:2211.15305 [hep-th]} \BibitemShut
	{NoStop}%
	\bibitem [{\citenamefont {Wang}\ \emph {et~al.}(2020)\citenamefont {Wang},
		\citenamefont {Sheng}, \citenamefont {Lu}, \citenamefont {Gao}, \citenamefont
		{Chang}, \citenamefont {Pang}, \citenamefont {Yang}, \citenamefont {Zhu},
		\citenamefont {Liu},\ and\ \citenamefont {Jin}}]{Wang:2020ypl}%
	\BibitemOpen
	\bibfield  {author} {\bibinfo {author} {\bibfnamefont {Y.}~\bibnamefont
			{Wang}}, \bibinfo {author} {\bibfnamefont {C.}~\bibnamefont {Sheng}},
		\bibinfo {author} {\bibfnamefont {Y.-H.}\ \bibnamefont {Lu}}, \bibinfo
		{author} {\bibfnamefont {J.}~\bibnamefont {Gao}}, \bibinfo {author}
		{\bibfnamefont {Y.-J.}\ \bibnamefont {Chang}}, \bibinfo {author}
		{\bibfnamefont {X.-L.}\ \bibnamefont {Pang}}, \bibinfo {author}
		{\bibfnamefont {T.-H.}\ \bibnamefont {Yang}}, \bibinfo {author}
		{\bibfnamefont {S.-N.}\ \bibnamefont {Zhu}}, \bibinfo {author} {\bibfnamefont
			{H.}~\bibnamefont {Liu}},\ and\ \bibinfo {author} {\bibfnamefont {X.-M.}\
			\bibnamefont {Jin}},\ }\bibfield  {title} {\bibinfo {title} {{Quantum
				simulation of particle pair creation near the event horizon}},\ }\href
	{https://doi.org/10.1093/nsr/nwaa111} {\bibfield  {journal} {\bibinfo
			{journal} {Natl. Sci. Rev.}\ }\textbf {\bibinfo {volume} {7}},\ \bibinfo
		{pages} {1476} (\bibinfo {year} {2020})}\BibitemShut {NoStop}%
	\bibitem [{\citenamefont {Koke}\ \emph {et~al.}(2016)\citenamefont {Koke},
		\citenamefont {Noh},\ and\ \citenamefont {Angelakis}}]{Koke:2016etw}%
	\BibitemOpen
	\bibfield  {author} {\bibinfo {author} {\bibfnamefont {C.}~\bibnamefont
			{Koke}}, \bibinfo {author} {\bibfnamefont {C.}~\bibnamefont {Noh}},\ and\
		\bibinfo {author} {\bibfnamefont {D.~G.}\ \bibnamefont {Angelakis}},\
	}\bibfield  {title} {\bibinfo {title} {{Dirac equation in 2-dimensional
				curved spacetime, particle creation, and coupled waveguide arrays}},\ }\href
	{https://doi.org/10.1016/j.aop.2016.08.013} {\bibfield  {journal} {\bibinfo
			{journal} {Annals Phys.}\ }\textbf {\bibinfo {volume} {374}},\ \bibinfo
		{pages} {162} (\bibinfo {year} {2016})},\ \Eprint
	{https://arxiv.org/abs/1607.04821} {arXiv:1607.04821 [quant-ph]} \BibitemShut
	{NoStop}%
	\bibitem [{\citenamefont {Toga}\ \emph {et~al.}(2025)\citenamefont {Toga},
		\citenamefont {Samlodia},\ and\ \citenamefont {Kemper}}]{Toga2025FastSI}%
	\BibitemOpen
	\bibfield  {author} {\bibinfo {author} {\bibfnamefont {G.~C.}\ \bibnamefont
			{Toga}}, \bibinfo {author} {\bibfnamefont {A.}~\bibnamefont {Samlodia}},\
		and\ \bibinfo {author} {\bibfnamefont {A.~F.}\ \bibnamefont {Kemper}},\
	}\bibfield  {title} {\bibinfo {title} {Fast scrambling in the hyperbolic
			ising model}\ }(\bibinfo {year} {2025})\BibitemShut {NoStop}%
	\bibitem [{\citenamefont {Gong}\ and\ \citenamefont
		{Yang}(2025)}]{Gong2025DigitQS}%
	\BibitemOpen
	\bibfield  {author} {\bibinfo {author} {\bibfnamefont {J.-Q.}\ \bibnamefont
			{Gong}}\ and\ \bibinfo {author} {\bibfnamefont {J.-C.}\ \bibnamefont
			{Yang}},\ }\bibfield  {title} {\bibinfo {title} {Digit quantum simulation of
			a fermion field in an expanding universe}\ }(\bibinfo {year}
	{2025})\BibitemShut {NoStop}%
	\bibitem [{\citenamefont {Alkac}\ and\ \citenamefont
		{{\"O}zg{\"u}n}(2025)}]{Alkac:2025hrv}%
	\BibitemOpen
	\bibfield  {author} {\bibinfo {author} {\bibfnamefont {G.}~\bibnamefont
			{Alkac}}\ and\ \bibinfo {author} {\bibfnamefont {E.}~\bibnamefont
			{{\"O}zg{\"u}n}},\ }\bibfield  {title} {\bibinfo {title} {{Simulating Hawking
				radiation in quantum many-body systems: Deviations from the thermal
				spectrum}},\ }\href {https://doi.org/10.1016/j.physletb.2025.139783}
	{\bibfield  {journal} {\bibinfo  {journal} {Phys. Lett. B}\ }\textbf
		{\bibinfo {volume} {868}},\ \bibinfo {pages} {139783} (\bibinfo {year}
		{2025})},\ \Eprint {https://arxiv.org/abs/2502.08199} {arXiv:2502.08199
		[gr-qc]} \BibitemShut {NoStop}%
	\bibitem [{\citenamefont {Li}\ \emph {et~al.}(2024)\citenamefont {Li},
		\citenamefont {Li}, \citenamefont {Zhuang},\ and\ \citenamefont
		{Yung}}]{Li2024SimulatingTS}%
	\BibitemOpen
	\bibfield  {author} {\bibinfo {author} {\bibfnamefont {X.-W.}\ \bibnamefont
			{Li}}, \bibinfo {author} {\bibfnamefont {F.}~\bibnamefont {Li}}, \bibinfo
		{author} {\bibfnamefont {J.}~\bibnamefont {Zhuang}},\ and\ \bibinfo {author}
		{\bibfnamefont {M.-H.}\ \bibnamefont {Yung}},\ }\bibfield  {title} {\bibinfo
		{title} {Simulating the schwinger model with a regularized variational
			quantum imaginary time evolution}\ }(\bibinfo {year} {2024})\BibitemShut
	{NoStop}%
	\bibitem [{\citenamefont {Horner}\ \emph {et~al.}(2023)\citenamefont {Horner},
		\citenamefont {Hallam},\ and\ \citenamefont {Pachos}}]{Horner:2022sei}%
	\BibitemOpen
	\bibfield  {author} {\bibinfo {author} {\bibfnamefont {M.~D.}\ \bibnamefont
			{Horner}}, \bibinfo {author} {\bibfnamefont {A.}~\bibnamefont {Hallam}},\
		and\ \bibinfo {author} {\bibfnamefont {J.~K.}\ \bibnamefont {Pachos}},\
	}\bibfield  {title} {\bibinfo {title} {{Chiral Spin-Chain Interfaces
				Exhibiting Event-Horizon Physics}},\ }\href
	{https://doi.org/10.1103/PhysRevLett.130.016701} {\bibfield  {journal}
		{\bibinfo  {journal} {Phys. Rev. Lett.}\ }\textbf {\bibinfo {volume} {130}},\
		\bibinfo {pages} {016701} (\bibinfo {year} {2023})},\ \Eprint
	{https://arxiv.org/abs/2207.08840} {arXiv:2207.08840 [cond-mat.str-el]}
	\BibitemShut {NoStop}%
	\bibitem [{\citenamefont {Kinoshita}\ \emph
		{et~al.}(2025{\natexlab{b}})\citenamefont {Kinoshita}, \citenamefont
		{Murata}, \citenamefont {Yamamoto},\ and\ \citenamefont
		{Yoshii}}]{Kinoshita2025SpinSA}%
	\BibitemOpen
	\bibfield  {author} {\bibinfo {author} {\bibfnamefont {S.}~\bibnamefont
			{Kinoshita}}, \bibinfo {author} {\bibfnamefont {K.}~\bibnamefont {Murata}},
		\bibinfo {author} {\bibfnamefont {D.}~\bibnamefont {Yamamoto}},\ and\
		\bibinfo {author} {\bibfnamefont {R.}~\bibnamefont {Yoshii}},\ }\bibfield
	{title} {\bibinfo {title} {Spin systems as quantum field theories in
			inflationary universe: A study with unruh-dewitt detectors}\ }(\bibinfo
	{year} {2025})\BibitemShut {NoStop}%
	\bibitem [{\citenamefont {Daniel}\ \emph
		{et~al.}(2025{\natexlab{a}})\citenamefont {Daniel}, \citenamefont {Hallam},
		\citenamefont {Horner},\ and\ \citenamefont {Pachos}}]{Daniel:2024boy}%
	\BibitemOpen
	\bibfield  {author} {\bibinfo {author} {\bibfnamefont {A.}~\bibnamefont
			{Daniel}}, \bibinfo {author} {\bibfnamefont {A.}~\bibnamefont {Hallam}},
		\bibinfo {author} {\bibfnamefont {M.~D.}\ \bibnamefont {Horner}},\ and\
		\bibinfo {author} {\bibfnamefont {J.~K.}\ \bibnamefont {Pachos}},\ }\bibfield
	{title} {\bibinfo {title} {{Optimally scrambling chiral spin-chain with
				effective black hole geometry}},\ }\href
	{https://doi.org/10.1038/s41598-025-92760-9} {\bibfield  {journal} {\bibinfo
			{journal} {Sci. Rep.}\ }\textbf {\bibinfo {volume} {15}},\ \bibinfo {pages}
		{9103} (\bibinfo {year} {2025}{\natexlab{a}})},\ \Eprint
	{https://arxiv.org/abs/2404.14473} {arXiv:2404.14473 [cond-mat.str-el]}
	\BibitemShut {NoStop}%
	\bibitem [{\citenamefont {Jaiswal}\ and\ \citenamefont
		{Shankaranarayanan}(2025)}]{Jaiswal:2025euo}%
	\BibitemOpen
	\bibfield  {author} {\bibinfo {author} {\bibfnamefont {N.}~\bibnamefont
			{Jaiswal}}\ and\ \bibinfo {author} {\bibfnamefont {S.}~\bibnamefont
			{Shankaranarayanan}},\ }\bibfield  {title} {\bibinfo {title} {{Analog charged
				black hole formation via percolation: Exploring cosmic censorship and Hoop
				conjecture}},\ }\href {https://doi.org/10.1103/PhysRevD.111.L101502}
	{\bibfield  {journal} {\bibinfo  {journal} {Phys. Rev. D}\ }\textbf {\bibinfo
			{volume} {111}},\ \bibinfo {pages} {L101502} (\bibinfo {year} {2025})},\
	\Eprint {https://arxiv.org/abs/2502.09317} {arXiv:2502.09317 [gr-qc]}
	\BibitemShut {NoStop}%
	\bibitem [{\citenamefont {Moghaddam}\ \emph {et~al.}(2025)\citenamefont
		{Moghaddam}, \citenamefont {Konye}, \citenamefont {Mertens}, \citenamefont
		{van Wezel},\ and\ \citenamefont {van~den Brink}}]{Moghaddam2025SyntheticHA}%
	\BibitemOpen
	\bibfield  {author} {\bibinfo {author} {\bibfnamefont {A.~G.}\ \bibnamefont
			{Moghaddam}}, \bibinfo {author} {\bibfnamefont {V.}~\bibnamefont {Konye}},
		\bibinfo {author} {\bibfnamefont {L.}~\bibnamefont {Mertens}}, \bibinfo
		{author} {\bibfnamefont {J.}~\bibnamefont {van Wezel}},\ and\ \bibinfo
		{author} {\bibfnamefont {J.}~\bibnamefont {van~den Brink}},\ }\bibfield
	{title} {\bibinfo {title} {Synthetic horizons and thermalization in an atomic
			chain and its relation to quantum hall systems}\ }(\bibinfo {year}
	{2025})\BibitemShut {NoStop}%
	\bibitem [{\citenamefont {Daniel}\ \emph
		{et~al.}(2025{\natexlab{b}})\citenamefont {Daniel}, \citenamefont {Bhore},
		\citenamefont {Pachos}, \citenamefont {Liu},\ and\ \citenamefont
		{Hallam}}]{Daniel2025QuantumTB}%
	\BibitemOpen
	\bibfield  {author} {\bibinfo {author} {\bibfnamefont {A.}~\bibnamefont
			{Daniel}}, \bibinfo {author} {\bibfnamefont {T.}~\bibnamefont {Bhore}},
		\bibinfo {author} {\bibfnamefont {J.~K.}\ \bibnamefont {Pachos}}, \bibinfo
		{author} {\bibfnamefont {C.}~\bibnamefont {Liu}},\ and\ \bibinfo {author}
		{\bibfnamefont {A.}~\bibnamefont {Hallam}},\ }\bibfield  {title} {\bibinfo
		{title} {Quantum teleportation between simulated binary black holes}\
	}(\bibinfo {year} {2025})\BibitemShut {NoStop}%
	\bibitem [{\citenamefont {Barouch}\ \emph {et~al.}(1970)\citenamefont
		{Barouch}, \citenamefont {McCoy},\ and\ \citenamefont
		{Dresden}}]{Barouch:1970ryz}%
	\BibitemOpen
	\bibfield  {author} {\bibinfo {author} {\bibfnamefont {E.}~\bibnamefont
			{Barouch}}, \bibinfo {author} {\bibfnamefont {B.~M.}\ \bibnamefont {McCoy}},\
		and\ \bibinfo {author} {\bibfnamefont {M.}~\bibnamefont {Dresden}},\
	}\bibfield  {title} {\bibinfo {title} {{Statistical Mechanics of the XY
				Model. I}},\ }\href {https://doi.org/10.1103/PhysRevA.2.1075} {\bibfield
		{journal} {\bibinfo  {journal} {Phys. Rev. A}\ }\textbf {\bibinfo {volume}
			{2}},\ \bibinfo {pages} {1075} (\bibinfo {year} {1970})}\BibitemShut
	{NoStop}%
	\bibitem [{\citenamefont {Barouch}\ and\ \citenamefont
		{McCoy}(1971)}]{Barouch:1971ywx}%
	\BibitemOpen
	\bibfield  {author} {\bibinfo {author} {\bibfnamefont {E.}~\bibnamefont
			{Barouch}}\ and\ \bibinfo {author} {\bibfnamefont {B.~M.}\ \bibnamefont
			{McCoy}},\ }\bibfield  {title} {\bibinfo {title} {{Statistical Mechanics of
				the XY Model. II. Spin-Correlation Functions}},\ }\href
	{https://doi.org/10.1103/PhysRevA.3.786} {\bibfield  {journal} {\bibinfo
			{journal} {Phys. Rev. A}\ }\textbf {\bibinfo {volume} {3}},\ \bibinfo {pages}
		{786} (\bibinfo {year} {1971})}\BibitemShut {NoStop}%
	\bibitem [{\citenamefont {Garc\'\i{}a-Mata}\ \emph {et~al.}(2023)\citenamefont
		{Garc\'\i{}a-Mata}, \citenamefont {Jalabert},\ and\ \citenamefont
		{Wisniacki}}]{Garcia-Mata:2022voo}%
	\BibitemOpen
	\bibfield  {author} {\bibinfo {author} {\bibfnamefont {I.}~\bibnamefont
			{Garc\'\i{}a-Mata}}, \bibinfo {author} {\bibfnamefont {R.~A.}\ \bibnamefont
			{Jalabert}},\ and\ \bibinfo {author} {\bibfnamefont {D.~A.}\ \bibnamefont
			{Wisniacki}},\ }\bibfield  {title} {\bibinfo {title} {{Out-of-time-order
				correlators and quantum chaos}},\ }\href
	{https://doi.org/10.4249/scholarpedia.55237} {\bibfield  {journal} {\bibinfo
			{journal} {Scholarpedia}\ }\textbf {\bibinfo {volume} {18}},\ \bibinfo
		{pages} {55237} (\bibinfo {year} {2023})},\ \Eprint
	{https://arxiv.org/abs/2209.07965} {arXiv:2209.07965 [quant-ph]} \BibitemShut
	{NoStop}%
	\bibitem [{\citenamefont {Roberts}\ and\ \citenamefont
		{Swingle}(2016)}]{PhysRevLett.117.091602}%
	\BibitemOpen
	\bibfield  {author} {\bibinfo {author} {\bibfnamefont {D.~A.}\ \bibnamefont
			{Roberts}}\ and\ \bibinfo {author} {\bibfnamefont {B.}~\bibnamefont
			{Swingle}},\ }\bibfield  {title} {\bibinfo {title} {Lieb-robinson bound and
			the butterfly effect in quantum field theories},\ }\href
	{https://doi.org/10.1103/PhysRevLett.117.091602} {\bibfield  {journal}
		{\bibinfo  {journal} {Phys. Rev. Lett.}\ }\textbf {\bibinfo {volume} {117}},\
		\bibinfo {pages} {091602} (\bibinfo {year} {2016})}\BibitemShut {NoStop}%
	\bibitem [{\citenamefont {Tian}\ \emph {et~al.}(2022)\citenamefont {Tian},
		\citenamefont {Lin}, \citenamefont {Fischer},\ and\ \citenamefont
		{Du}}]{Tian:2020bze}%
	\BibitemOpen
	\bibfield  {author} {\bibinfo {author} {\bibfnamefont {Z.}~\bibnamefont
			{Tian}}, \bibinfo {author} {\bibfnamefont {Y.}~\bibnamefont {Lin}}, \bibinfo
		{author} {\bibfnamefont {U.~R.}\ \bibnamefont {Fischer}},\ and\ \bibinfo
		{author} {\bibfnamefont {J.}~\bibnamefont {Du}},\ }\bibfield  {title}
	{\bibinfo {title} {{Testing the upper bound on the speed of scrambling with
				an analogue of Hawking radiation using trapped ions}},\ }\href
	{https://doi.org/10.1140/epjc/s10052-022-10149-8} {\bibfield  {journal}
		{\bibinfo  {journal} {Eur. Phys. J. C}\ }\textbf {\bibinfo {volume} {82}},\
		\bibinfo {pages} {212} (\bibinfo {year} {2022})},\ \Eprint
	{https://arxiv.org/abs/2007.05949} {arXiv:2007.05949 [quant-ph]} \BibitemShut
	{NoStop}%
	\bibitem [{\citenamefont {Wei}\ \emph {et~al.}(2018)\citenamefont {Wei},
		\citenamefont {Ramanathan},\ and\ \citenamefont {Cappellaro}}]{Wei:2016jjg}%
	\BibitemOpen
	\bibfield  {author} {\bibinfo {author} {\bibfnamefont {K.~X.}\ \bibnamefont
			{Wei}}, \bibinfo {author} {\bibfnamefont {C.}~\bibnamefont {Ramanathan}},\
		and\ \bibinfo {author} {\bibfnamefont {P.}~\bibnamefont {Cappellaro}},\
	}\bibfield  {title} {\bibinfo {title} {{Exploring Localization in Nuclear
				Spin Chains}},\ }\href {https://doi.org/10.1103/PhysRevLett.120.070501}
	{\bibfield  {journal} {\bibinfo  {journal} {Phys. Rev. Lett.}\ }\textbf
		{\bibinfo {volume} {120}},\ \bibinfo {pages} {070501} (\bibinfo {year}
		{2018})},\ \Eprint {https://arxiv.org/abs/1612.05249} {arXiv:1612.05249
		[cond-mat.dis-nn]} \BibitemShut {NoStop}%
	\bibitem [{\citenamefont {Wang}\ \emph {et~al.}(2025)\citenamefont {Wang} \emph
		{et~al.}}]{Wang:2025ade}%
	\BibitemOpen
	\bibfield  {author} {\bibinfo {author} {\bibfnamefont {Z.~T.}\ \bibnamefont
			{Wang}} \emph {et~al.},\ }\bibfield  {title} {\bibinfo {title} {{Observing
				Two-Particle Correlation Dynamics in Tunable Superconducting Bose-Hubbard
				Simulators}},\ }\href@noop {} {\  (\bibinfo {year} {2025})},\ \Eprint
	{https://arxiv.org/abs/2509.02180} {arXiv:2509.02180 [quant-ph]} \BibitemShut
	{NoStop}%
	\bibitem [{\citenamefont {Li}\ \emph {et~al.}(2025)\citenamefont {Li} \emph
		{et~al.}}]{Li:2025kje}%
	\BibitemOpen
	\bibfield  {author} {\bibinfo {author} {\bibfnamefont {T.-M.}\ \bibnamefont
			{Li}} \emph {et~al.},\ }\bibfield  {title} {\bibinfo {title} {{Many-body
				delocalization with a two-dimensional 70-qubit superconducting quantum
				simulator}},\ }\href@noop {} {\  (\bibinfo {year} {2025})},\ \Eprint
	{https://arxiv.org/abs/2507.16882} {arXiv:2507.16882 [quant-ph]} \BibitemShut
	{NoStop}%
	\bibitem [{\citenamefont {Zhao}\ \emph
		{et~al.}(2025{\natexlab{b}})\citenamefont {Zhao} \emph
		{et~al.}}]{Zhao:2025jcc}%
	\BibitemOpen
	\bibfield  {author} {\bibinfo {author} {\bibfnamefont {K.}~\bibnamefont
			{Zhao}} \emph {et~al.},\ }\bibfield  {title} {\bibinfo {title} {{Microwave
				engineering of tunable spin interactions with superconducting qubits}},\
	}\href {https://doi.org/10.1063/5.0281890} {\bibfield  {journal} {\bibinfo
			{journal} {Appl. Phys. Lett.}\ }\textbf {\bibinfo {volume} {127}},\ \bibinfo
		{pages} {064001} (\bibinfo {year} {2025}{\natexlab{b}})},\ \Eprint
	{https://arxiv.org/abs/2505.16286} {arXiv:2505.16286 [quant-ph]} \BibitemShut
	{NoStop}%
	\bibitem [{\citenamefont {Liu}\ \emph {et~al.}(2025{\natexlab{c}})\citenamefont
		{Liu} \emph {et~al.}}]{Liu:2025tys}%
	\BibitemOpen
	\bibfield  {author} {\bibinfo {author} {\bibfnamefont {Z.-H.}\ \bibnamefont
			{Liu}} \emph {et~al.},\ }\bibfield  {title} {\bibinfo {title}
		{{Prethermalization by Random Multipolar Driving on a 78-Qubit
				Superconducting Processor}},\ }\href@noop {} {\  (\bibinfo {year}
		{2025}{\natexlab{c}})},\ \Eprint {https://arxiv.org/abs/2503.21553}
	{arXiv:2503.21553 [quant-ph]} \BibitemShut {NoStop}%
	\bibitem [{\citenamefont {Alam}\ \emph {et~al.}(2025)\citenamefont {Alam} \emph
		{et~al.}}]{Alam:2025ofn}%
	\BibitemOpen
	\bibfield  {author} {\bibinfo {author} {\bibfnamefont {F.}~\bibnamefont
			{Alam}} \emph {et~al.},\ }\bibfield  {title} {\bibinfo {title} {{Programmable
				digital quantum simulation of 2D Fermi-Hubbard dynamics using 72
				superconducting qubits}},\ }\href@noop {} {\  (\bibinfo {year} {2025})},\
	\Eprint {https://arxiv.org/abs/2510.26845} {arXiv:2510.26845 [quant-ph]}
	\BibitemShut {NoStop}%
	\bibitem [{\citenamefont {Yan}\ \emph {et~al.}(2018)\citenamefont {Yan},
		\citenamefont {Krantz}, \citenamefont {Sung}, \citenamefont {Kjaergaard},
		\citenamefont {Campbell}, \citenamefont {Orlando}, \citenamefont
		{Gustavsson},\ and\ \citenamefont {Oliver}}]{Yan:2018mli}%
	\BibitemOpen
	\bibfield  {author} {\bibinfo {author} {\bibfnamefont {F.}~\bibnamefont
			{Yan}}, \bibinfo {author} {\bibfnamefont {P.}~\bibnamefont {Krantz}},
		\bibinfo {author} {\bibfnamefont {Y.}~\bibnamefont {Sung}}, \bibinfo {author}
		{\bibfnamefont {M.}~\bibnamefont {Kjaergaard}}, \bibinfo {author}
		{\bibfnamefont {D.~L.}\ \bibnamefont {Campbell}}, \bibinfo {author}
		{\bibfnamefont {T.~P.}\ \bibnamefont {Orlando}}, \bibinfo {author}
		{\bibfnamefont {S.}~\bibnamefont {Gustavsson}},\ and\ \bibinfo {author}
		{\bibfnamefont {W.~D.}\ \bibnamefont {Oliver}},\ }\bibfield  {title}
	{\bibinfo {title} {{Tunable Coupling Scheme for Implementing High-Fidelity
				Two-Qubit Gates}},\ }\href {https://doi.org/10.1103/PhysRevApplied.10.054062}
	{\bibfield  {journal} {\bibinfo  {journal} {Phys. Rev. Appl.}\ }\textbf
		{\bibinfo {volume} {10}},\ \bibinfo {pages} {054062} (\bibinfo {year}
		{2018})},\ \Eprint {https://arxiv.org/abs/1803.09813} {arXiv:1803.09813
		[quant-ph]} \BibitemShut {NoStop}%
	\bibitem [{\citenamefont {Bravyi}\ \emph {et~al.}(2011)\citenamefont {Bravyi},
		\citenamefont {DiVincenzo},\ and\ \citenamefont {Loss}}]{Bravyi:2011fda}%
	\BibitemOpen
	\bibfield  {author} {\bibinfo {author} {\bibfnamefont {S.}~\bibnamefont
			{Bravyi}}, \bibinfo {author} {\bibfnamefont {D.~P.}\ \bibnamefont
			{DiVincenzo}},\ and\ \bibinfo {author} {\bibfnamefont {D.}~\bibnamefont
			{Loss}},\ }\bibfield  {title} {\bibinfo {title}
		{{Schrieffer{\textendash}Wolff transformation for quantum many-body
				systems}},\ }\href {https://doi.org/10.1016/j.aop.2011.06.004} {\bibfield
		{journal} {\bibinfo  {journal} {Annals Phys.}\ }\textbf {\bibinfo {volume}
			{326}},\ \bibinfo {pages} {2793} (\bibinfo {year} {2011})},\ \Eprint
	{https://arxiv.org/abs/1105.0675} {arXiv:1105.0675 [quant-ph]} \BibitemShut
	{NoStop}%
	\bibitem [{\citenamefont {Nieuwenhuizen}\ \emph {et~al.}(2004)\citenamefont
		{Nieuwenhuizen}, \citenamefont {Balian},\ and\ \citenamefont
		{Allahverdyan}}]{Nieuwenhuizen:2004exd}%
	\BibitemOpen
	\bibfield  {author} {\bibinfo {author} {\bibfnamefont {T.~M.}\ \bibnamefont
			{Nieuwenhuizen}}, \bibinfo {author} {\bibfnamefont {R.}~\bibnamefont
			{Balian}},\ and\ \bibinfo {author} {\bibfnamefont {A.~E.}\ \bibnamefont
			{Allahverdyan}},\ }\bibfield  {title} {\bibinfo {title} {{Maximal work
				extraction from finite quantum systems}},\ }\href
	{https://doi.org/10.1209/epl/i2004-10101-2} {\bibfield  {journal} {\bibinfo
			{journal} {EPL}\ }\textbf {\bibinfo {volume} {67}},\ \bibinfo {pages} {565}
		(\bibinfo {year} {2004})},\ \Eprint {https://arxiv.org/abs/cond-mat/0401574}
	{arXiv:cond-mat/0401574} \BibitemShut {NoStop}%
	\bibitem [{\citenamefont {Swingle}(2018)}]{Swingle:2018ekw}%
	\BibitemOpen
	\bibfield  {author} {\bibinfo {author} {\bibfnamefont {B.}~\bibnamefont
			{Swingle}},\ }\bibfield  {title} {\bibinfo {title} {{Unscrambling the physics
				of out-of-time-order correlators}},\ }\href
	{https://doi.org/10.1038/s41567-018-0295-5} {\bibfield  {journal} {\bibinfo
			{journal} {Nature Phys.}\ }\textbf {\bibinfo {volume} {14}},\ \bibinfo
		{pages} {988} (\bibinfo {year} {2018})}\BibitemShut {NoStop}%
\end{thebibliography}
%

\end{document}